\documentclass[fleqn,usenatbib]{mnras}

\usepackage{newtxtext,newtxmath}

\usepackage[T1]{fontenc}
\usepackage{ae,aecompl}
\usepackage[dvipsnames]{xcolor}
\usepackage[toc,page]{appendix}

\usepackage{graphicx}	
\usepackage{amsmath}	
\usepackage{booktabs}
\usepackage{longtable}

\newcommand{\GG}[1]{}

\title[The \textit{False Widow} Link]{The \textit{False Widow} Link Between Neutron Star X-ray Binaries and Spider Pulsars}

\author[A. H. Knight et al.]{
Amy H. Knight,$^{1}$\thanks{E-mail: amy.knight@physics.ox.ac.uk}
Adam Ingram,$^{2}$ 
Jakob van den Eijnden,$^{1}$
Douglas J. K. Buisson,$^{3}$
Lauren Rhodes,$^{1}$ \and 
\ and Matthew Middleton$^{4}$
\\
$^{1}$Department of Physics, Astrophysics, University of Oxford, Denys Wilkinson Building, Keble Road, Oxford, OX1 3RH, UK\\
$^{2}$School of Mathematics, Statistics, and Physics, Newcastle University, Newcastle upon Tyne NE1 7RU, UK\\
$^{3}$Independent\\
$^{4}$School of Physics and Astronomy, University of Southampton, Highfield, Southampton, SO17 1BJ, UK\\
}

\date{Accepted XXX. Received YYY; in original form ZZZ}

\pubyear{2021}

\begin{document}
\label{firstpage}
\pagerange{\pageref{firstpage}--\pageref{lastpage}}
\maketitle

\begin{abstract}
The discovery of transitional millisecond pulsars (tMSPs) provided conclusive proof that neutron star (NS) low-mass X-ray binaries (LMXBs) comprise part of the evolutionary pathway towards binary millisecond pulsars (MSPs). Redback and black widow `spider' pulsars are a sub-category of binary MSPs that `devour' their companions through ablation - the process through which material is lifted from the stellar surface by a pulsar wind. In addition to reducing the companion star's mass, ablation introduces observable characteristics like extended, energy-dependent and asymmetric eclipse profiles in systems observed at a sufficiently high inclination. Here, we present a detailed study and comparison of the X-ray eclipses of two NS LMXBs; \textit{Swift} J1858.6$-$0814 and EXO 0748$-$676. Some of the X-ray eclipse characteristics observed in these two LMXBs are similar to the radio eclipse characteristics of eclipsing redback and black widow pulsars, suggesting that they may also host ablated companion stars. X-ray irradiation or a pulsar wind could drive the ablation. We conduct orbital phase-resolved spectroscopy for both LMXBs to map the column density, ionization and covering fraction of the material outflow. From this, we infer the presence of highly ionized and clumpy ablated material around the companion star in both systems. We term LMXBs undergoing ablation, \textit{false widows}, and speculate that they may be the progenitors of redback pulsars under the assumption that ablation begins in the LMXB stage. Therefore, the false widows could provide a link between LMXBs and spider pulsars. The detection of radio pulsations during non-accreting states can support this hypothesis.
\end{abstract}




\section{Introduction}
Neutron stars (NSs) observed in various astrophysical environments prompt investigations into the possible evolutionary pathways that could result in that observed stage. One clear example is the discovery of the first radio pulsar \citep{Bell1967} and the first radio millisecond pulsar (MSP) \citep{Backer1982, Alpar1982} through to the detection of millisecond X-ray pulsations in SAX J1808.4-3658 \citep{Wijnands1998}, which provided the first concrete evidence that MSPs begin life as ordinary pulsars. As part of a binary system, the NS (pulsar) is spun-up to millisecond spin periods through the accretion of matter from its companion star. During this accretion-powered phase, the system exists as an X-ray binary (XRB) but may transition into a rotation-powered state when accretion ceases. Systems observed to switch between an accretion-powered state and a rotation-powered state are transitional millisecond pulsars (tMSPs) and relate XRBs to binary MSPs \citep{Archibald2009, Papitto2013}.

MSPs in binary systems can align with one of the three sub-classes of spider pulsars, depending on their properties. These sub-classes are redback, black widow and huntsman pulsars and are distinguished by the companion star's mass or classification or the binary's orbital period. Huntsman pulsars are categorised by orbital periods $\gtrsim 5$ days and sub-giant companions \citep{Swihart2018}. Redbacks host companion stars with masses $M_{\rm{cs}} \sim 0.2 - 0.4 M_{\odot}$, while black widows host less massive companion stars $M_{\rm{cs}} << 0.1 M_{\odot}$ \citep{Roberts2013a, Chen2013}. Redbacks and black widows both have short orbital periods of $\sim$ hours. Characteristically, the companion star in redback and black widow pulsars undergoes ablation driven by the pulsar wind and irradiation from the NS \citep{Kluzniak1988, Ruderman1989a, Ruderman1989b}. The pulsar wind and radiation incident on the companion star liberate its outer layers, giving rise to a collection of ablated material around the system. This so-called cannibalistic behaviour is where the naming of redbacks and black widows originated. Conversely huntsman pulsars, who take after their larger, non-aggressive namesake, do not experience ablation. 

Ablated material remains gravitationally bound to the binary, thus introducing characteristic observable properties like delayed pulse arrival times \citep{Polzin2018} and a material trail behind the companion star \citep{Fruchter1988}. Furthermore, sufficiently inclined redbacks and black widows can exhibit radio eclipses which are frequency-dependent, extended and asymmetric due to absorption by the ablated material \citep{Polzin2018, Polzin2020}. Ablation increases mass loss from the companion star. Therefore, redback and black widow pulsars may comprise part of the evolutionary pathway towards isolated MSPs. Given sufficient time, the ablation process could reduce the mass of a redback companion so that it classifies as a black widow. If ablation continues beyond this, the companion may be entirely \textit{devoured} by ablation, resulting in an isolated MSP. This scenario is similar to the dissolution or evaporation of the donor star, which is considered a likely scenario for the formation of isolated MSPs \citep{Bildsten2002}. In this case, the donor star must remain hot (e.g. irradiated) while losing mass to entirely evaporate, and the donor may expand in the process. However, ablation by a pulsar wind is not an efficient process \citep{Ginzburg2020}, so this alone cannot explain the very low mass companions observed within redback and black widow pulsars, nor can it be the sole mechanism responsible for the complete evaporation of companion stars in redback and black widow pulsars.

Despite ablation typically being a property of the rotation-powered redback and black widow pulsars, evidence has recently been uncovered, suggesting that it begins in the LMXB phase for short-period binaries \citep{Knight2022a, Knight2022b}
and could continue throughout the tMSP phase. Some redback pulsars are also tMSPs (see \citealt{Linares2014} for discussion of the transitional redback pulsars PSR J1023$+$0038, PSR J1824$-$24521 and XSS J12270$-$4859), so LMXBs hosting ablated companion stars could be the progenitors of transitional redback pulsars. TMSPs switch between accretion-powered X-ray pulsations and rotation-powered radio pulsations on timescales as short as weeks, implying the transitional phase is an extended stage of binary pulsar evolution, although the initial transition from LMXB to tMSP could take several years \citep{Archibald2009, Papitto2013, Bahramian2018}. During the initial and transitional accretion-powered (LMXB) phases, the companion star loses mass via Roche lobe overflow (RLOF) and X-ray-driven ablation. The combination of these processes reduces the mass of the companion, paving the way towards the under-massive companion stars observed within redback and black widow systems \citep{Fruchter1988, Stapper1996}, especially if the transitional phase is long-lived. 

Two eclipsing NS LXMBs, EXO 0748$-$676 and \textit{Swift} J1858.6$-$0814 (hereafter, EXO 0748 and Sw J1858 respectively), show evidence that their companions are undergoing X-ray driven ablation. EXO 0748 was discovered in 1985 with \textit{EXOSAT} and regularly monitored with \textit{RXTE} and \textit{XMM-Newton} during its $\gtrsim 20$ year outburst. The observed X-ray eclipses occur when the $\sim 0.4 M_{\odot}$ M-dwarf companion passes in front of the NS and accretion disc. The eclipses last $\sim 500$ s, recur on the orbital period of $3.824$ hrs and show highly variable transition durations \citep{Parmar1991, Wolff2009}. EXO 0748 entered X-ray quiescence in late 2009 (see \citealt{Degenaar2011} for a summary). In contrast, Sw J1858 was discovered in 2018 as an X-ray transient \citep{Krimm2018}, showing drastic flaring behaviour analogous to V404 Cyg and V4641 Sgr \citep{Wijnands2000, Revnivtsev2002, Walton2017, Motta2017}. It transitioned from this so-called flaring state to a more steady outburst phase in 2020 ($\sim$ MJD 58885; See in Fig. 1 of \citealt{Buisson2021}) during which type I X-ray bursts \citep{Buisson2020} and X-ray eclipses of $\sim 4100$ s duration were uncovered \citep{Buisson2021}. The orbital period was determined to be $\approx 21.3$ hrs \citep{Buisson2021}, and the companion was found to be an evolved, low mass star, $M_{\rm cs} \sim 0.3 M_{\odot}$, with a large radius, $R_{\rm cs} \sim 1.0 R_{\odot}$ \citep{Buisson2021,Knight2022b}. Sw J1858 entered X-ray quiescence in May 2020 \citep{Saikia2020}.

By modelling the X-ray eclipse profiles of EXO 0748, \citet{Knight2022a} were able to infer that a narrow region of ablated material surrounds the companion which extends $\sim 700 - 1500$ km from the stellar surface ($0.2 - 0.5\% R_{\rm{cs}}$). This material extends further from the companion on the egress side of the star than the ingress side, explaining the observed eclipse asymmetry. The ingress and egress durations are $15.2$s and $17.5$s, respectively \citep{Knight2022a}. Additionally, the material absorbs softer X-rays more efficiently than harder X-rays causing the observed eclipse profiles to depend on energy (see Fig. 2 of \citealt{Knight2022a}) and be extended in time. These observed features of the X-ray eclipses are analogous to the radio eclipse characteristics of redback and black widow pulsars \citep{Polzin2020}, although, it's important to note that in redbacks and black widows, the radio eclipses typically last a much larger fraction of the orbital period than the X-ray eclipses studied here. No X-ray or radio pulsations have currently been found in EXO 0748 (see \citealt{Jain2011} \footnote{\citet{Jain2011} searched for X-ray pulsations in a frequency range based upon the detection of burst oscillations at $45$ Hz by \citet{Villarreal_2004}. To date, there has not been a published X-ray pulsation search at the burst oscillation frequency of 552 Hz \citep{Galloway2010}, or a published search for radio pulsations at either frequency.}  and \citealt{Ratti2012}), but quiescent studies of the source uncovered spider-like features which led \citet{Ratti2012} and \citet{Parikh2020} to respectively describe EXO 0748 as `black widow-like' and `transitional redback-like'. The X-ray eclipses of Sw J1858 also share several characteristics with EXO 0748, redback and black widow pulsars. While determining the binary inclination ($i \sim 81^{\circ}$) and mass ratio ($q \sim 0.14$), \citet{Knight2022b} uncovered a highly extended, asymmetric ablated layer around the companion star that extends $7000 - 14000$ km from the stellar surface ($1 - 1.6 \% R_{\rm{cs}}$). The duration of the ingress and egress were determined to be $\sim 106$s and $\sim 174$s respectively, and the eclipse profiles displayed the same energy dependence as EXO 0748 (see Fig. 2 of \citealt{Knight2022b}). There are no observations of X-ray or radio pulsations from Sw J1858 at the time of writing.

In addition to the remarkably similar eclipse characteristics shared by both EXO 0748 and Sw J1858 (and by redback and black widow pulsars), they share several non-eclipse similarities including flares, stellar prominences during the ingress and out-of-eclipse dips which occur preferentially during pre-ingress phases \citep{Wolff2009,Buisson2021}. As such, we group EXO 0748 and Sw J1858 together and refer to them as \textit{false widows}, defined as LMXBs in which the companion star is undergoing ablation. As arachnids, the less aggressive false widows are often mistakenly identified as black widows due to their visual similarity. Thus, the naming of these binaries as false widows refers to the key observable similarity of ablation, but implies a less ablated companion. In addition to a detailed comparison of the previously uncovered eclipse characteristics of Sw J1858 \citep{Knight2022b} and EXO 0748 \citep{Knight2022a}, we present a phase-resolved spectral analysis of the near-eclipse epochs of both systems to determine the extent and influence of the ablated material. We also show how the eclipse transition durations change over time and infer the role of the ablated material on these changes. 

In Section \ref{sx:Data} we present the data used in this work and the reduction procedures. We present the phase-resolved spectral analysis in Section \ref{sx:PRS} and investigate the evolution of the eclipse transition durations in Section \ref{sx:Transitions}. In Section \ref{sx:Discuss}, we thoroughly compare the two systems and speculate how \textit{false widows} relate to spider pulsars. We present our conclusions in Section \ref{sx:Conclude}.  

\section{Data Reduction}
\label{sx:Data}
\begin{figure*}
\centering
\includegraphics[width=\textwidth]{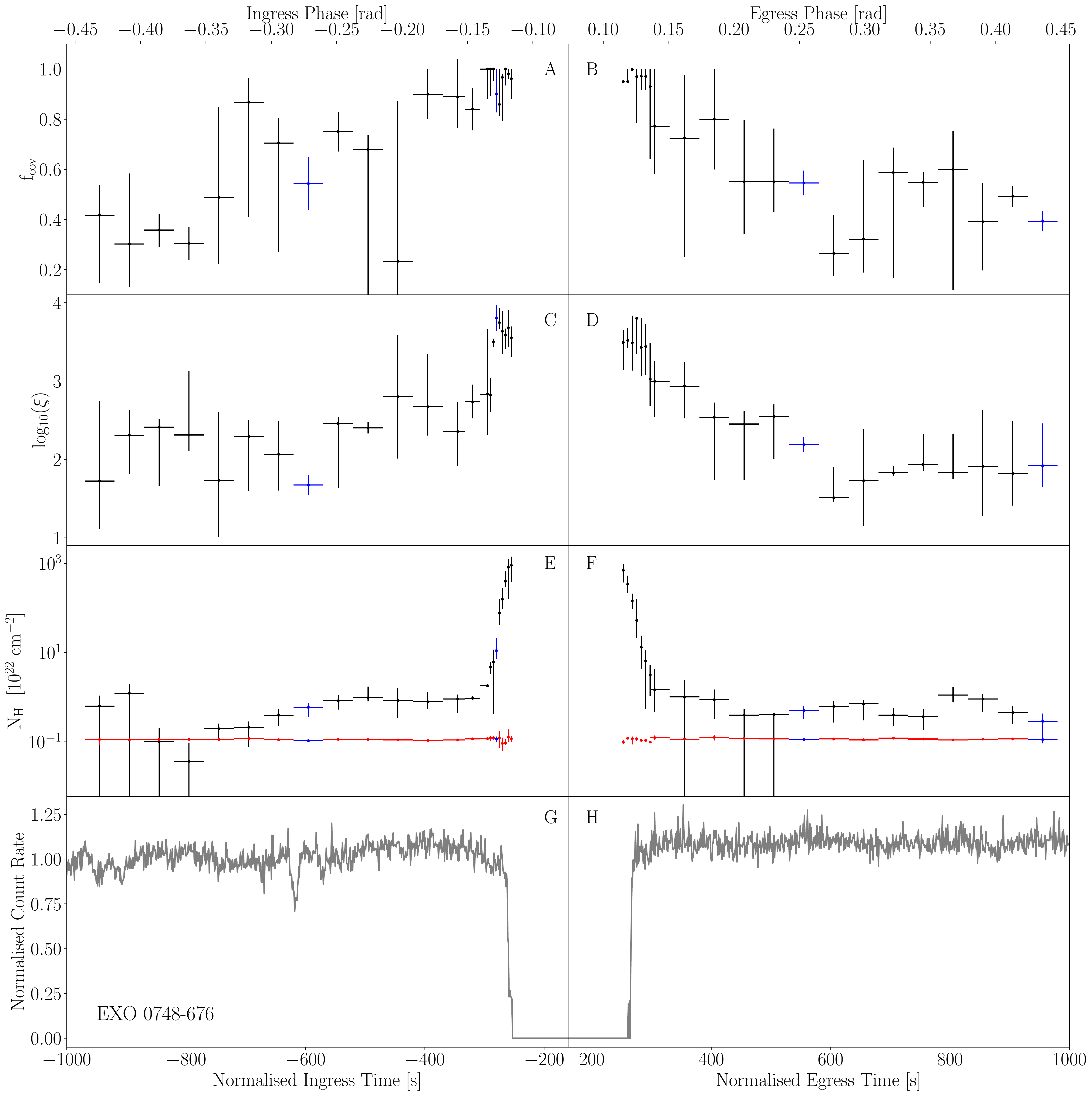}
\vspace{-0.5cm}
\caption{Phase resolved spectroscopy of the near-eclipse epochs in EXO 0748 showing how the covering fraction (A and B), log of the ionization parameter defined as $\log(L/(\rm{n}R^{2}))$ (C and D) and column density (E and F) vary as a function of orbital phase. Here the red data points show the measured column density of the ISM. The corresponding \textit{XMM-Newton} eclipse profile is shown in panels G and H. A total of 44 spectra are extracted from 6 non-overlapping energy bands: $0.2 - 0.5$ keV, $0.5 - 1.0$ keV, $1.0 - 2.0$ keV, $2.0 - 4.0$ keV, $4.0 - 6.0$ keV and $6.0 - 8.0$ keV. The ingress and egress spectra are extracted with time bins of $2.5$ s and $5.0$s respectively, while the pre-ingress and post-egress spectra were both extracted with $50.0$ s time bins. The spectra are modelled with our local \textsc{xspec} model, \textsc{abssca} \citep{Knight2022a} using cash-statistics (c-stat) due to the low count rate in several of the spectra. Error bars are $1 \sigma$. The blue data points are used to highlight the representative spectra analysed in detail in Appendix \ref{sx:prs}.
}
\label{fig:EXO_PRS}
\end{figure*}

\begin{figure*}
\centering
\includegraphics[width=\textwidth]{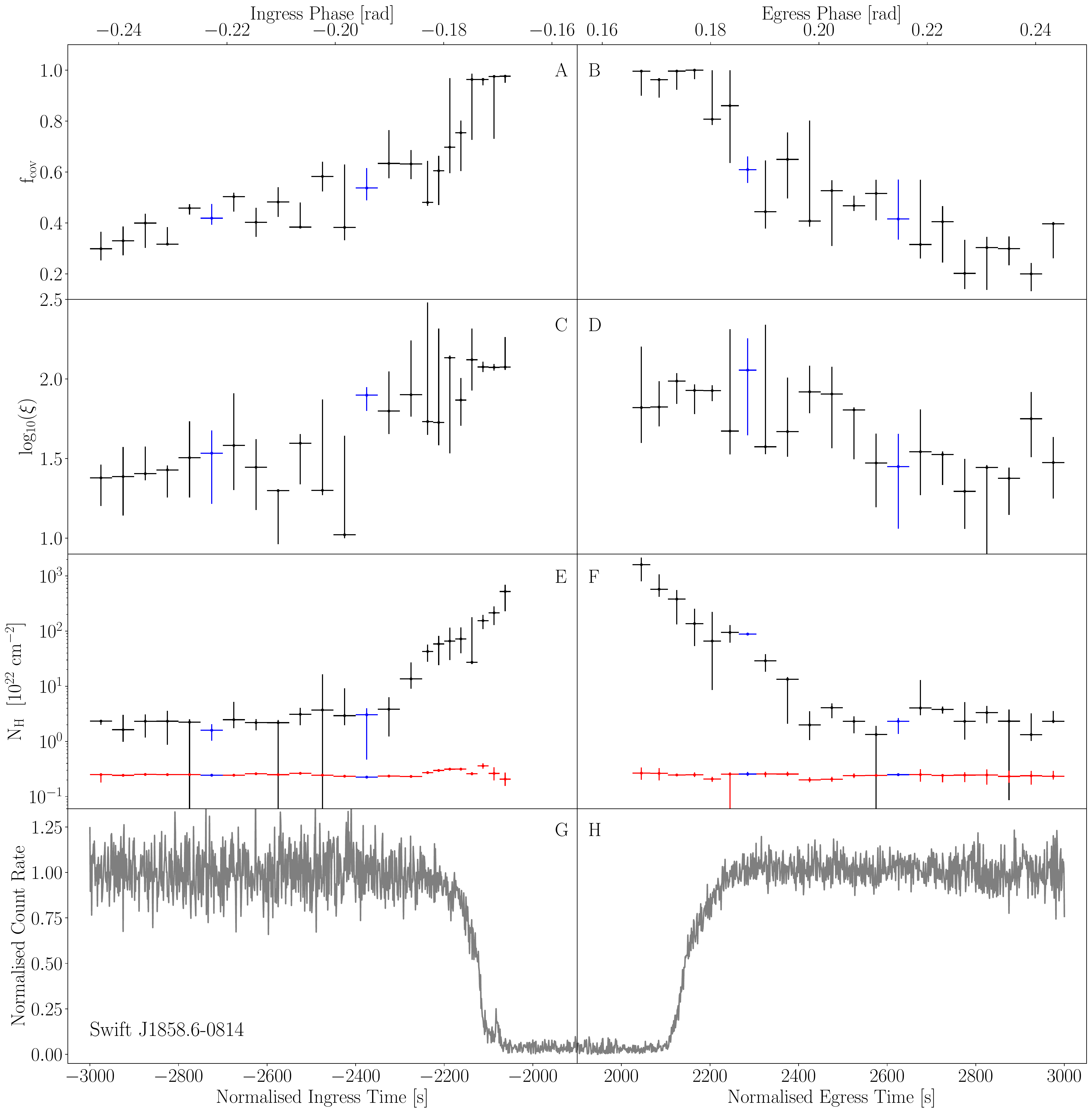}
\vspace{-0.5cm}
\caption{Phase resolved spectroscopy of the near-eclipse regions in Sw J1858 showing how the covering fraction (A and B), log of the ionization parameter defined as $\log(L/(\rm{n}R^{2}))$ (C and D) and column density (E and F) vary as a function of orbital phase. Here the red data points show the measured column density of the ISM. The corresponding \textit{NICER} eclipse profile is shown in panels G and H. A total of 44 spectra are extracted from 6 non-overlapping energy bands: $0.5 - 1.0$ keV, $1.0 - 2.0$ keV, $2.0 - 4.0$ keV, $4.0 - 6.0$ keV, $6.0 - 8.0$ keV and $8.0 - 10.0$ keV. The ingress and egress spectra are extracted with time bins of $25.0$ s and $40.0$s respectively, while the pre-ingress and post-egress spectra were both extracted with $50.0$ s time bins. The spectra are modelled with our local \textsc{xspec} model, \textsc{abssca} \citep{Knight2022a} using cash-statistics (c-stat) due to the low count rate in several of the spectra. Error bars are $1 \sigma$. The blue data points are used to highlight the representative spectra analysed in detail in Appendix \ref{sx:prs}.
}
\label{fig:J1858_PRS}
\end{figure*}

\begin{figure}
\centering
\includegraphics[width=\columnwidth]{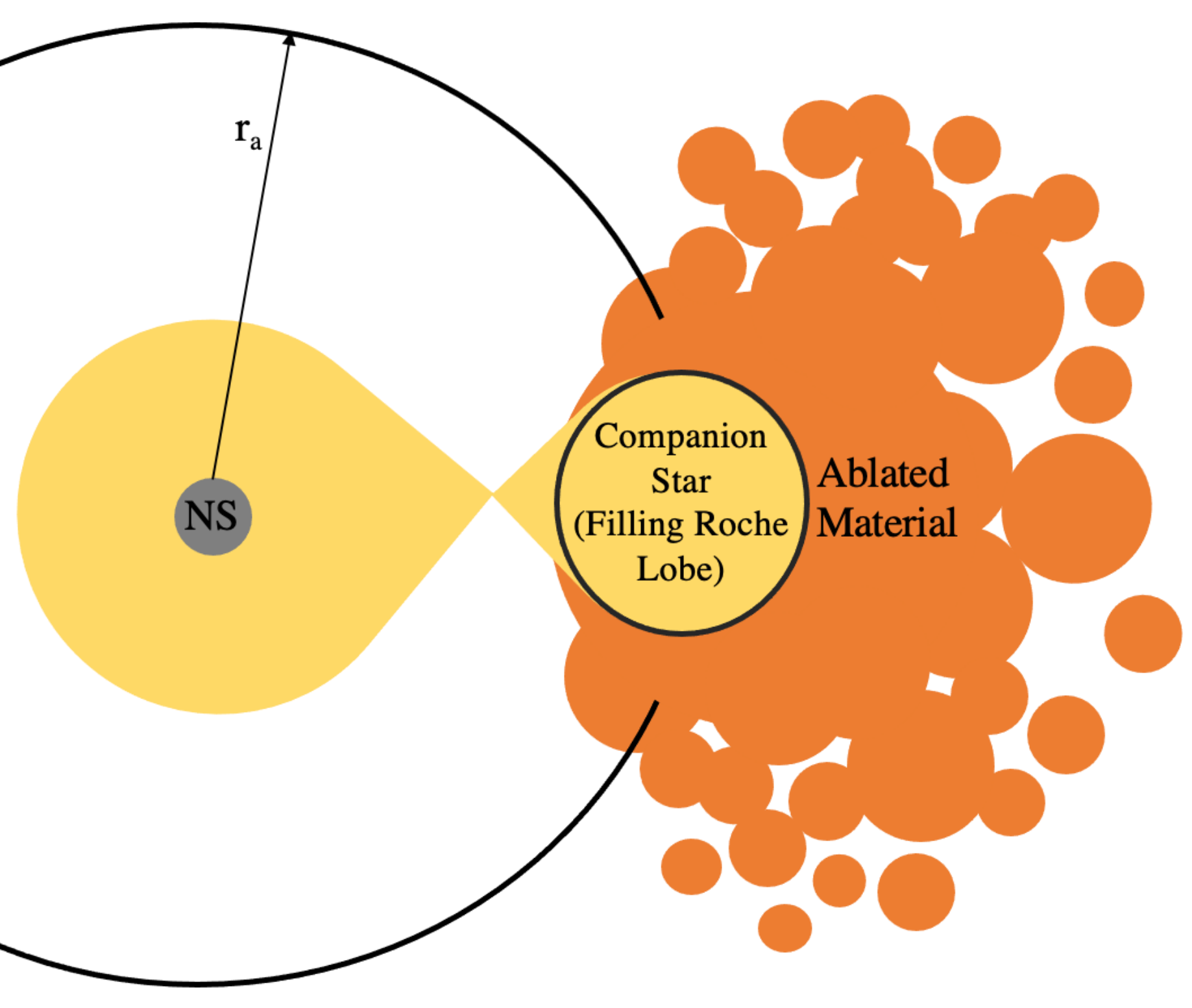}
\vspace{-0.5cm}
\caption{A schematic showing the geometry of a \textit{false widow} undergoing ablation. The NS and companion star have an assumed circular orbit with separation, $r_{\rm{a}}$, and the companion star is expected to be filling its Roche lobe. The ablated material is liberated from the companion star by incident X-rays from the NS and accretion disc. The material clumps near to the companion star and then diffuses away such that the density of the material remains high but the overall absorbing column decreases. Note, this schematic is not to scale.
}
\label{fig:Ablation_Schematic}
\end{figure}

Three data sets are considered in this work. For Sw J1858 we consider all available archival \textit{NICER} observations of the source between November 2018 and July 2020. For EXO 0748, we consider all archival \textit{RXTE} observations that contain a full eclipse profile (429 out of 746 observations) and a single soft-state outburst observation taken by \textit{XMM-Newton} in April 2005 (Obs-ID 0212480501) \citep{Ponti2014}, which contains four eclipses. As we have previously utilised the \textit{NICER} and \textit{XMM-Newton} data to model the eclipse profiles of each source, the data reduction procedures are detailed in \citet{Knight2022b} and \citet{Knight2022a} respectively. The \textit{RXTE} data reduction pipeline is described below.

\subsection{\textit{RXTE} Pipeline}
Due to the large number of \textit{RXTE} observations considered in this work, we did not manually calibrate the data and extract light curves for each observation containing an eclipse. Instead, we applied the fully automated \textsc{chromos} pipeline\footnote{\url{https://github.com/davidgardenier/chromos}} \citep{Gardenier2018}: an X-ray spectral-timing analysis code that, for a given list of \textit{RXTE} ObsIDs, downloads the raw observational data from \textsc{HEASARC}, applies all necessary data reduction steps (depending on the observation mode) and extracts light curves at the native time resolution of the data mode in a band made up of energy channels most closely matching the user-defined energy range (which accounts for the changes to the \textit{RXTE} channel-to-energy conversion throughout its lifetime). In addition, \textsc{chromos} is capable of calculating more advanced spectral-timing observables, such as power colours and hardnesses. However, given the scope of this work, we run the pipeline up to the light curve extraction. We extract light curves in several different energy bands: $3-6$ keV, $6-10$ keV, $10-16$ keV and $2-15$ keV. All analysed ObsIDs are listed in Table \ref{tb:RXTE_Table}, including the number of active proportional counter units (PCUs), average $2-15$ keV count rate and hardness, defined as the ratio of the $10-16$ keV to the $6-10$ keV count rate. We note that, while the same energy bands were used for each ObsID, slight differences between the exact energy range may exist between observations due to shifts in the conversion between energy channel and photon energy over time, as well as differences in observing mode. The light curves are re-binned into $1$s time bins and normalised by dividing through by the mean out-of-eclipse count rate (such that the mean out-of-eclipse level is $1.0$). This normalisation typically results in a mean totality level of $0.2$ due to an in-eclipse count rate contributed from the background. 

\vspace*{-0.4cm}
\section{Phase-Resolved Spectroscopy}
\label{sx:PRS}
We investigate the near-eclipse epochs of the two false widows using phase-resolved spectroscopy to infer how the covering fraction, ionization and column density vary with orbital phase, $\phi$. 

For each source, we extract a total of 44 spectra; 23 spectra for where $\phi < 0$ for the pre-ingress and ingress spectra, and 21 spectra where $\phi > 0$ for the egress and post-egress spectra. Here, $\phi = 0$ corresponds to the phase at the centre of totality. For EXO 0748 we use \textit{XMM-Newton} spectra, of  count rate vs. energy,  extracted using 6 energy bands: $0.2 - 0.5$ keV, $0.5 - 1.0$ keV, $1.0 - 2.0$ keV, $2.0 - 4.0$ keV, $4.0 - 6.0$ keV and $6.0 - 8.0$ keV. The ingress and egress spectra are extracted with times bins of $2.5$ s and $5.0$s respectively, while the pre-ingress and post-egress spectra were both extracted with $50.0$ s time bins. The use of finer time bins during the eclipse transitions is necessary to understand the evolution of material properties during these short times and the difference in bin width accounts for the eclipse asymmetry \citep[e.g.][]{Knight2022a}. Similarly, for Sw J1858, we extract \textit{NICER} spectra (count rate vs. energy) using 6 energy bands: $0.5 - 1.0$ keV, $1.0 - 2.0$ keV, $2.0 - 4.0$ keV, $4.0 - 6.0$ keV, $6.0 - 8.0$ and $8.0 - 10.0$ keV. The ingress and egress spectra are extracted with $25.0$ s and $40.0$ s time bins respectively. Here, the larger time bins reflect the longer ingress and egress durations in Sw J1858 than EXO 0748 \citep{Buisson2021, Knight2022b}. The pre-ingress and post-egress spectra for Sw J1858 are both extracted with $50.0$ s time bins. Note that the use of 6 wide energy bands in each case ensures a reasonable number of counts per energy bin in the spectra and, for this reason, is preferable to finer energy bins. Despite this, the count rates in several of the spectra are still too low to carry out the modelling using $\chi^2$ statistics. We instead use Cash statistics \citep{Cash1979} (the C-stat option within \textsc{xpsec}), which is more appropriate for low count rate data (see Appendix B of \citealt{Arnaud1996}). We obtain $1 \sigma$ error contours using the \texttt{steppar} command within \textsc{xspec} assuming an appropriate range around the best fitting values, divided into 400 steps (see Appendix \ref{sx:prs} for full details).

The spectra are modelled using our previously published local \textsc{xspec} model called \textsc{abssca} \citep{Knight2022a}, which considers both the absorption and scattering of X-ray photons by the ablated material. In this model, the total transmitted specific intensity is given as 
\begin{equation}
    I_{\rm E} = I^0_{\rm E} \bigg\{ \rm{f}_{\rm cov} \exp\{-N_{\rm H} [\sigma(\textit{E}) + (n_{\rm e}/n_{\rm H})\sigma_{\rm T} ] \} + 1 - \rm{f}_{\rm cov} \bigg\},
\end{equation}
where $I^0_{\rm E}$ is the out-of-eclipse specific intensity determined respectively for both EXO 0748 and Sw J1858 by using the out-of-eclipse spectral fits from \citet{Knight2022a} and \citet{Knight2022b}. The covering fraction is $\rm{f}_{\rm cov}$, $\rm{N}_H$ is the material hydrogen column density, $\sigma_{\rm {T}}$ is the Thomson electron scattering cross-section and $\sigma(E)$ is the absorption cross-section, which is calculated using the \textsc{xspec} model \textsc{zxipcf} \citep{Miller2006} and depends on the ionization parameter, $\xi = L / (n R^2)$, where $L$ is the luminosity of the irradiating source, $n$ is the (number) density of the irradiated material and $R$ is the distance between the source and the irradiated material. The ratio of free electrons to hydrogen nuclei in the material is $\rm{n_e/n_H} = 1.0$ and the redshift is $z=0$. Note \textsc{abssca} is insensitive to the ratio $\rm{n_e/n_H}$ (see \citealt{Knight2022a} for further details). We also separate absorption arising due to the ionized, ablated material from absorption by the interstellar medium (ISM). We model ISM absorption with the \textsc{xspec} model \textsc{tbabs}, assuming the abundances of \citet{Wilms2000}. Due to the observed asymmetry in the eclipse profiles of both false widow candidates, we allow $\rm{f}_{\rm cov}$, $\rm{N}_{H}$ and $\log(\xi)$ to vary across all spectra. Note that this approach differs from our previously published phase-resolved spectroscopy of the ingress and egress of EXO 0748, in which we tied $\log(\xi)$ and $\rm{f}_{\rm cov}$ across the spectra \citep{Knight2022a}. This was appropriate since $\log(\xi)$ and $\rm{f}_{\rm cov}$ are mostly consistent within errors during the ingress and egress for EXO 0748, and we were only analysing the eclipse transitions. However, here we are modelling spectra across a much wider range of orbital phases, during which $\log(\xi)$ and $\rm{f}_{\rm cov}$ may change significantly. Therefore, we allow all parameters to vary in this analysis. This does introduce some additional modelling complications, including partial degeneracies between the parameters, which we explore in detail in Appendix \ref{sx:prs}. 

The results are respectively shown in Fig. \ref{fig:EXO_PRS} and Fig. \ref{fig:J1858_PRS} for EXO 0748 and Sw J1858, where we see some similarities. For both sources, the covering fraction (Figs. \ref{fig:EXO_PRS}A and \ref{fig:J1858_PRS}A), ionization parameter (Figs. \ref{fig:EXO_PRS}C and \ref{fig:J1858_PRS}C) and column density (Figs. \ref{fig:EXO_PRS}E and \ref{fig:J1858_PRS}E, black) of the ionized absorber generally increase with proximity to totality during the ingress phases, whereas the column density of ISM absorption (Figs. \ref{fig:EXO_PRS}E and \ref{fig:J1858_PRS}E, red) remains constant within uncertainties. These parameter behaviours are roughly mirrored during the egress phases (Figs. \ref{fig:EXO_PRS}B, D and F and \ref{fig:J1858_PRS}B, D and F).

Our interpretation of these results is demonstrated schematically in Fig. \ref{fig:Ablation_Schematic}. The increase in column density with proximity to the eclipse indicates that the density of the ablated material reduces with distance from the companion, as is expected from simple mass conservation considerations. However, the detection of an ionized absorber for orbital phases beyond the eclipse transitions ($|\phi| \gtrsim 0.2$ rad) indicates that some ablated material is still present far from the companion. The lower covering fraction further from eclipse indicates that the material breaks up into small clumps as it gets further from the companion. Clumpy outflows have been invoked for Vela X-1 and Cyg X-1, both high-mass XRBs, to explain features including variable absorption along the orbit \citep{Hanke2008, Grinberg2015, Grinberg2017, Diez2022}. In these cases, clumps arise from perturbations in the stellar wind from the OB companion stars, which fragment due to strong shocks in the wind, and variations in temperature and density \citep{Grinberg2017}. While typically associated with OB stars, radiation-driven outflows may also become clumpy far from low-mass stars \citep{Lagae2021} which could explain our findings.

The evolution of the ionization parameter with orbital phase implies that material further away from the companion star is also further away from the NS surface (i.e. the material is further from the source of ionising X-rays), as is pictured in Fig. \ref{fig:Ablation_Schematic}. Thus it may appear parabolic in shape like the pulsar wind-driven ablation in spider pulsars. Such a geometry is required to explain how the higher density material closer to the companion star has a higher ionization parameter than the lower density material far from the companion. If we were to assume that the density is simply proportional to the measured column density, $n \propto N_{\rm{H}}$, then this increase in distance from the NS would need to be dramatic. For example, the increase in $N_{\rm{H}}$ by a factor of $\sim 1000$ in our EXO 0748 fits would require an increase in distance by a factor of $\sim \sqrt{1000} \approx 30$. This seems unlikely to be the case, since the binary motion should cause material far from the centre of mass to become approximately uniform in phase around the binary. However, the column density also depends on the path length of our sight line through the material and on the covering fraction. Our fits indicate that the covering fraction is f$_{\rm cov} \sim 0.5$ far from the companion, and it is natural to assume that the material layer reduces in width with distance from the stellar surface (as is illustrated in Fig. \ref{fig:Ablation_Schematic}), meaning that the density drops off less dramatically than would otherwise be inferred. For example, assuming the path length reduces by a factor of $4$ far from the companion means that the distance from the NS only needs to increase by a factor $\sim 4$. In addition, these changes will have an effect on the filling factor, particularly if the material forms clumps as less of the line of sight will contain absorbing material, thus allowing material with a higher density and a lower $\xi$ for a fixed column and overall size. Finally, our fits indicate that the ablated material is asymmetric; the decrease in $\rm{N}_{\rm{H}}$ after the eclipse is longer in duration than the increase before the eclipse. The asymmetry is most likely due to the orbital motion of the binary systems (see \citealt{Knight2022a} and \citealt{Knight2022b}), therefore, this is a strong indication that ablated material exists external to the Roche lobe. We refer to this as super-Roche lobe material.

\section{Eclipse Analysis}
\label{sx:Transitions}
\subsection{Initial Duration Evolution}

\begin{figure}
\centering
\includegraphics[width=\columnwidth]{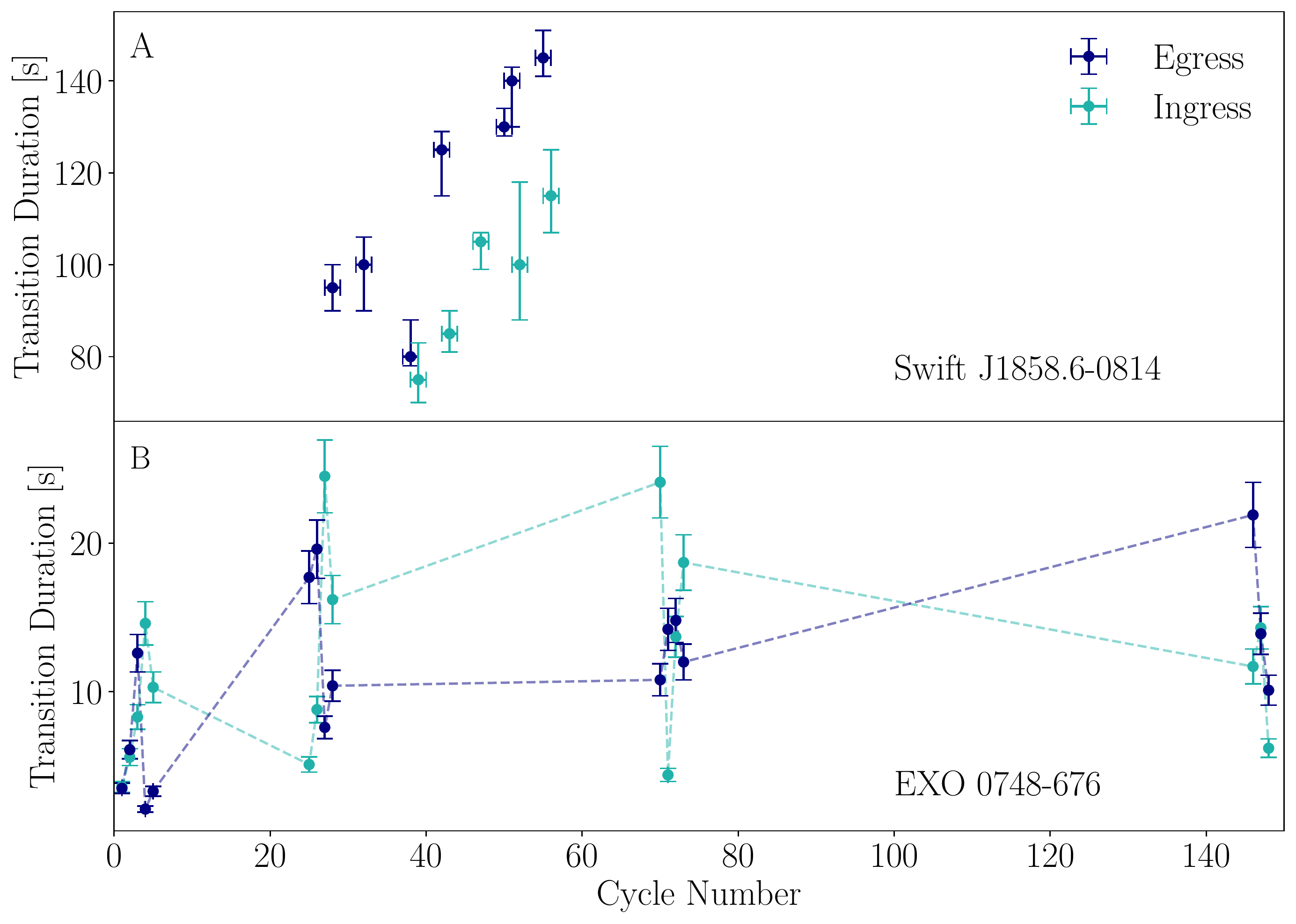}
\vspace{-0.5cm}
\caption{Measured eclipse transition durations for eclipses observed within the first 150 orbital cycles of the persistent outburst emission for Sw J1858 (panel A, \textit{NICER} observations) and EXO 0748 (panel B, \textit{EXOSAT} observations, \citealt{Parmar1991}). In both panels, the light and dark blue points represent the ingress and egress durations respectively. While both Sw J1858 and EXO 0748 show increasing transition durations in this stage, EXO 0748 shows very drastic variations attributed to an X-ray heated evaporating wind \citep{Parmar1991}.}
\label{fig:Early_J1858}
\end{figure}

After the onset of a persistent outburst, the ablated material layer is expected to expand until an equilibrium state is reached. During this time, the eclipse transition durations, are expected to increase as a result of increased line-of-sight absorption by the growing ablated layer. Following \cite{Knight2022b} and \cite{Knight2022a}, we measure the transition durations by identifying the times t$_{90}$ and t$_{10}$ as the times when the count rate first passes 90 and 10 percent respectively of the mean out-of-eclipse count rate (and stays above/below that level for at least 5 seconds; \citealt{Knight2022b}). Under this definition, the ingress starts at t$_{90}$ and ends at t$_{10}$ before totality, and the egress starts at t$_{10}$ and ends at t$_{90}$ after totality. Fig. \ref{fig:Early_J1858}A shows that the ingress and egress durations of Sw J1858 measured in this way did indeed increase as a function of orbital cycle at the start of the outburst. Here the first orbital cycle of each source's persistent outburst is defined as the zeroth cycle (e.g. \citealt{Knight2022b}). For Sw J1858, which showed a flaring outburst state prior to its persistent outburst state, this occurs at $\sim$ MJD 58885 \citep{Buisson2021}.

The onset of the persistent emission from EXO 0748 was observed by \textit{EXOSAT}. The eclipse transition durations measured from the eclipses observed by \textit{EXOSAT} between February 1985 and January 1986 were reported by \citet{Parmar1991}, who noted their drastic variations and, in some cases, long durations. \citet{Parmar1991} suggested that a cause of this behaviour could be an X-ray heated evaporating wind (ablated stellar material) that enhances the atmospheric scale height of the companion, extending the transition durations. These measurements are shown in Fig. \ref{fig:Early_J1858}B for the first 150 orbital cycles of EXO 0748, enabling comparison between the early eclipse transitions of EXO 0748 and Sw J1858. Due to the small number of observed eclipses of Sw J1858 and large variations in the transition durations of the EXO 0748 eclipses, making comparisons between the sources is not straightforward, however, both do show increases in transition duration on short timescales. As Sw J1858 is now in quiescence \citep{Saikia2020, Parikh2020a}, we do not know if similar variations to EXO 0748 would have been observed. Also, for this reason, and because of the limited number of observed eclipses of Sw J1858, the analysis presented in the remainder of this section was only possible for EXO 0748.

\subsection{Evolution of the Eclipse Transitions}
\begin{figure}
\centering
\includegraphics[width=\columnwidth]{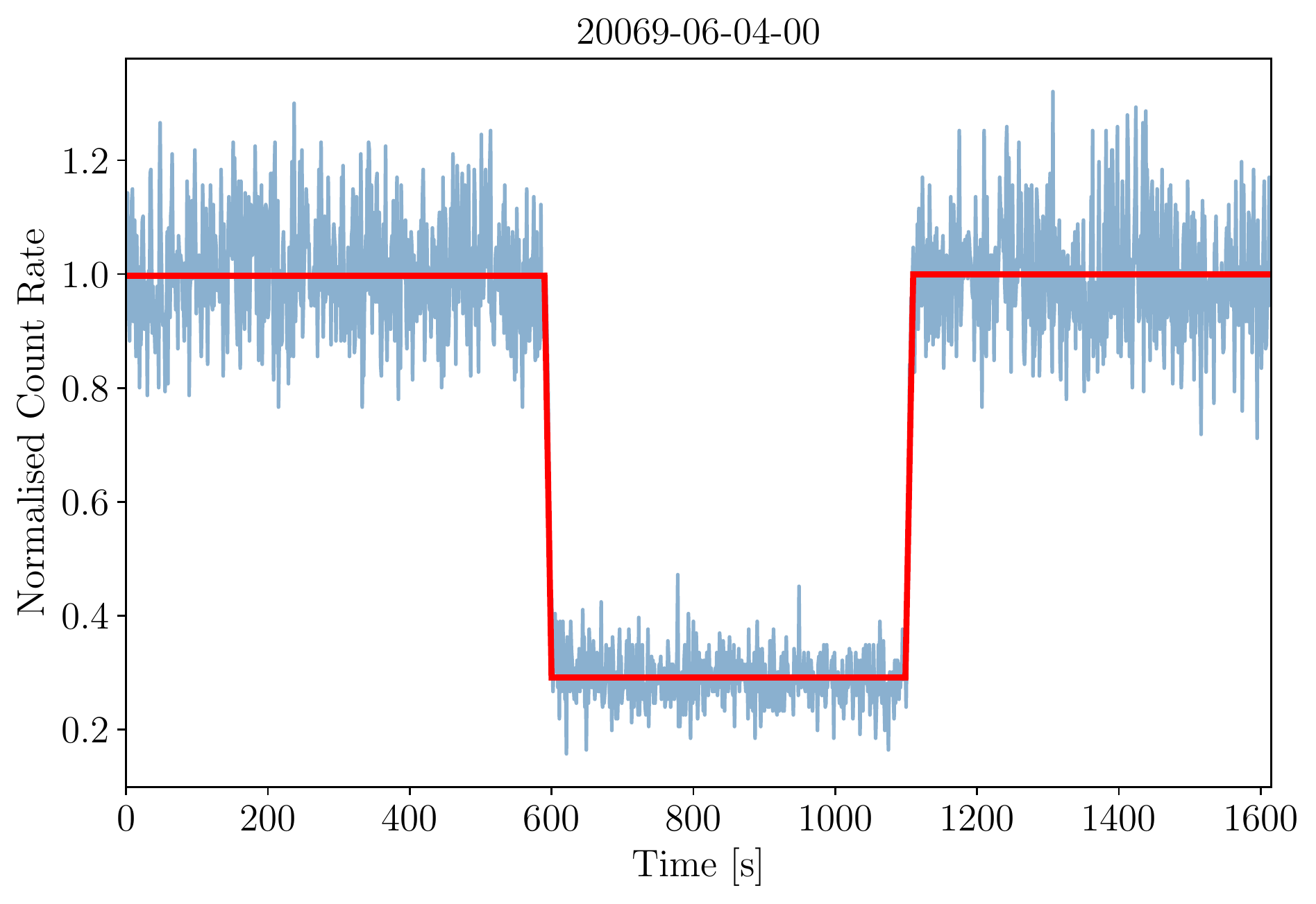}
\vspace{-0.5cm}
\caption{Example of a typical eclipse profile fit (red) obtained when fitting the simple eclipse profile model described in Section \ref{sx:Transitions} to the $2-15$ keV \textit{RXTE} light curves (blue) shown for ObsID 20069-06-04-00.}
\label{fig:ECFit}
\end{figure}

\begin{figure*}
\centering
\includegraphics[width=\textwidth]{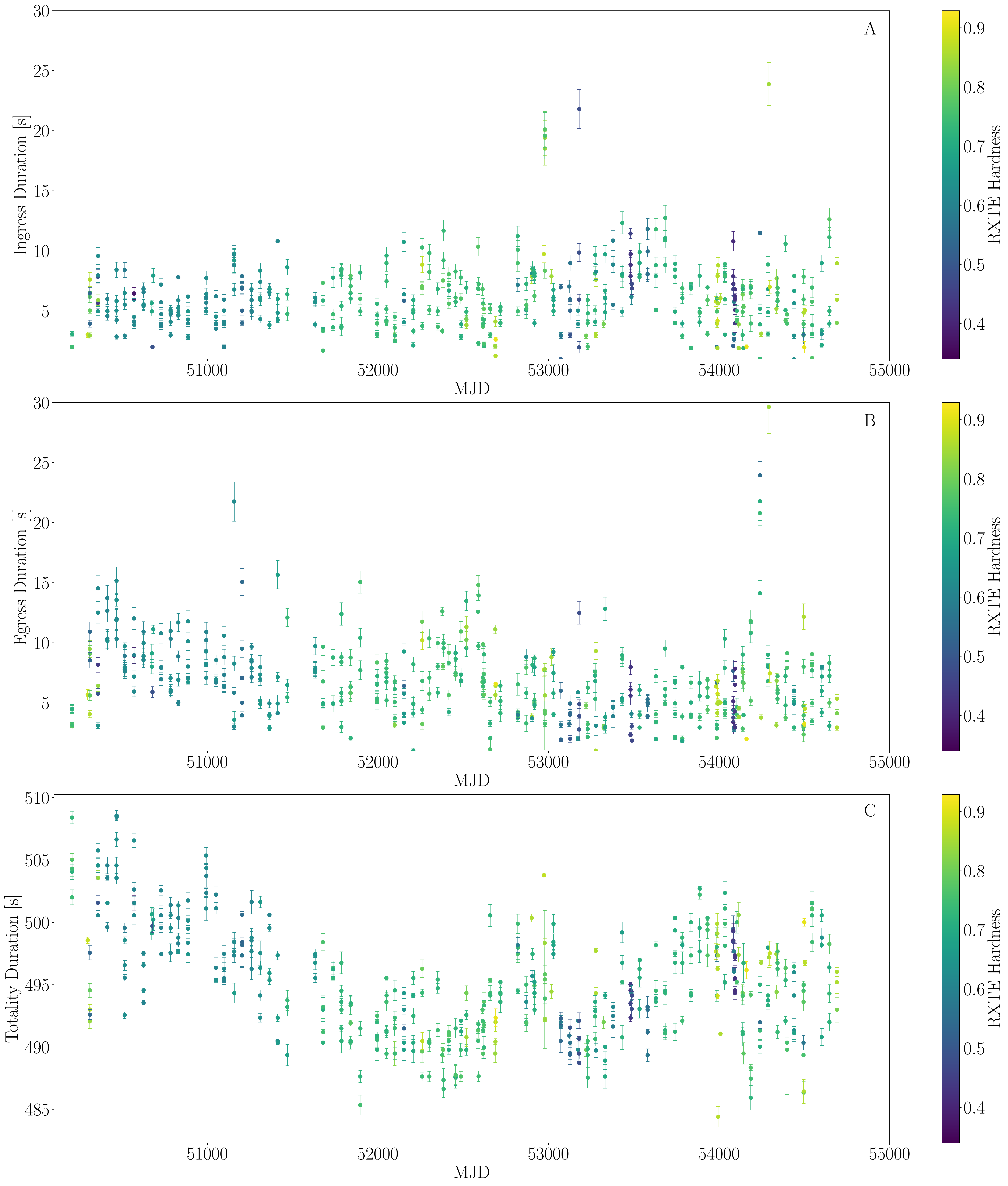}
\vspace{-0.5cm}
\caption{Ingress duration (panel A), egress duration (panel B) and totality duration (panel C) as functions of mid-eclipse MJD for all archival \textit{RXTE} observations of EXO 0748 that contain a full eclipse profile. The durations are measured by fitting the simple eclipse profile model, described in Section \ref{sx:Transitions}, to the $2-15$ keV eclipse profiles. Here each data point represents one ObsID. The variations observed in the eclipse transition times are indicative of ablated material clumping or changing structure as a result of the binary's orbital motion. The colour scale represents the \textit{RXTE} hardness, defined as ${F_{10-16 \rm{keV}}}/{F_{6-10 \rm{keV}}}$.}
\label{fig:RXTE_Times}
\end{figure*}

\begin{figure*}
\centering
\includegraphics[width=\textwidth]{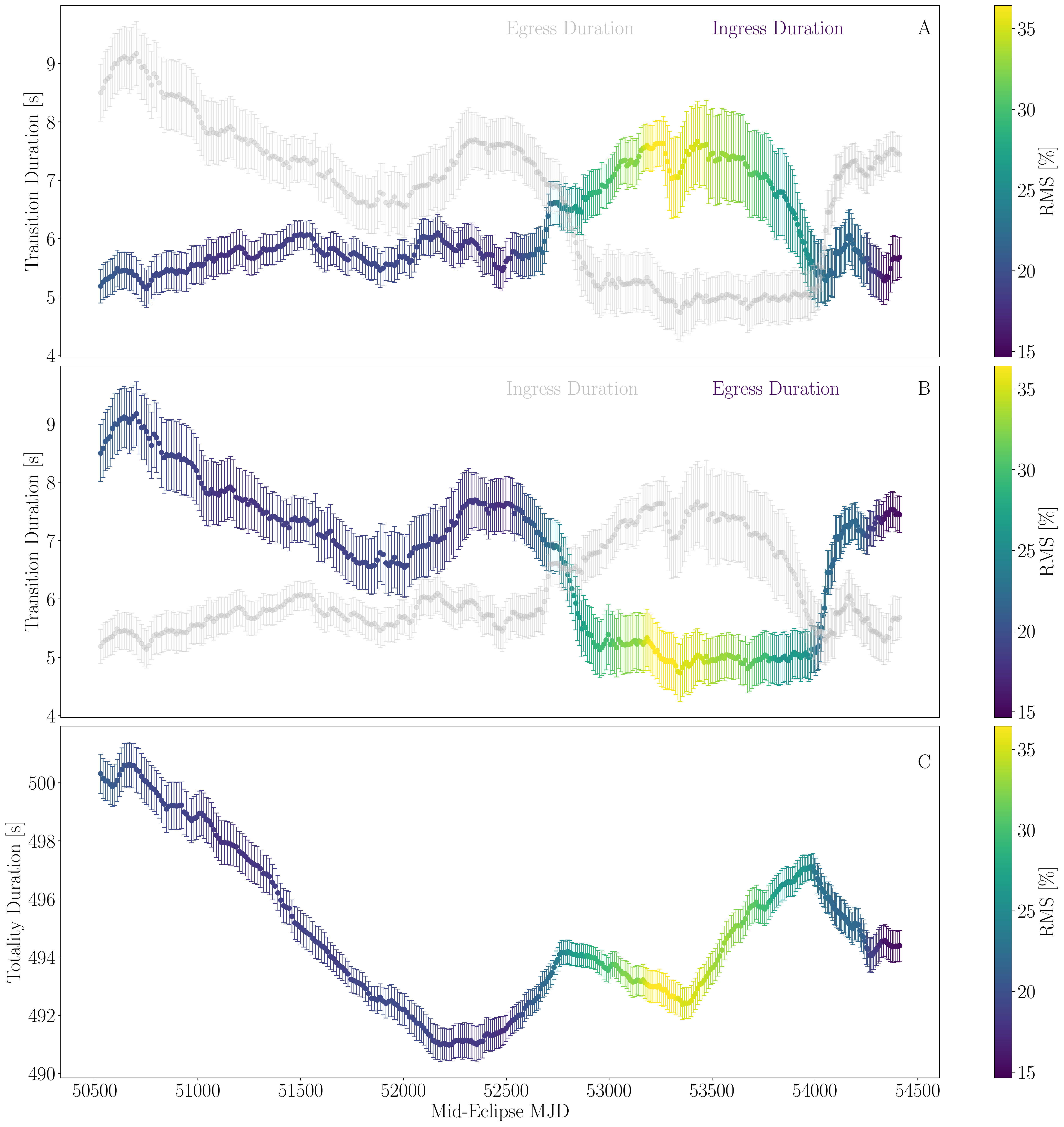}
\vspace{-0.5cm}
\caption{45-point moving averages of eclipse transition durations (Panels A and B) and totality duration (Panel C) as a functions mid-eclipse MJD and of background corrected root mean squared (RMS) variability amplitude (coloured). In Panels A and B respectively, the ingress and egress durations are coloured while the egress and ingress durations are over plot in grey for comparison, clearly showing the observed eclipse asymmetry and anti-correlation between the ingress and egress durations. The egress is typically longer in duration than the ingress, but reverses between $\sim 53000 - 54000$ MJD and seemingly coincides with an increase in RMS (transition to a harder state). In Panel C, the behaviour of the totality duration is determined to arise from changes to the orbital period of EXO 0748. In each panel, error bars correspond to the standard error, $\sigma / \sqrt{n}$ in each 45-point bin. Note that since this a moving average the error bars are not independent.
}
\label{fig:RXTE_Av_Times}
\end{figure*}

\begin{figure*}
\centering
\includegraphics[width=\textwidth]{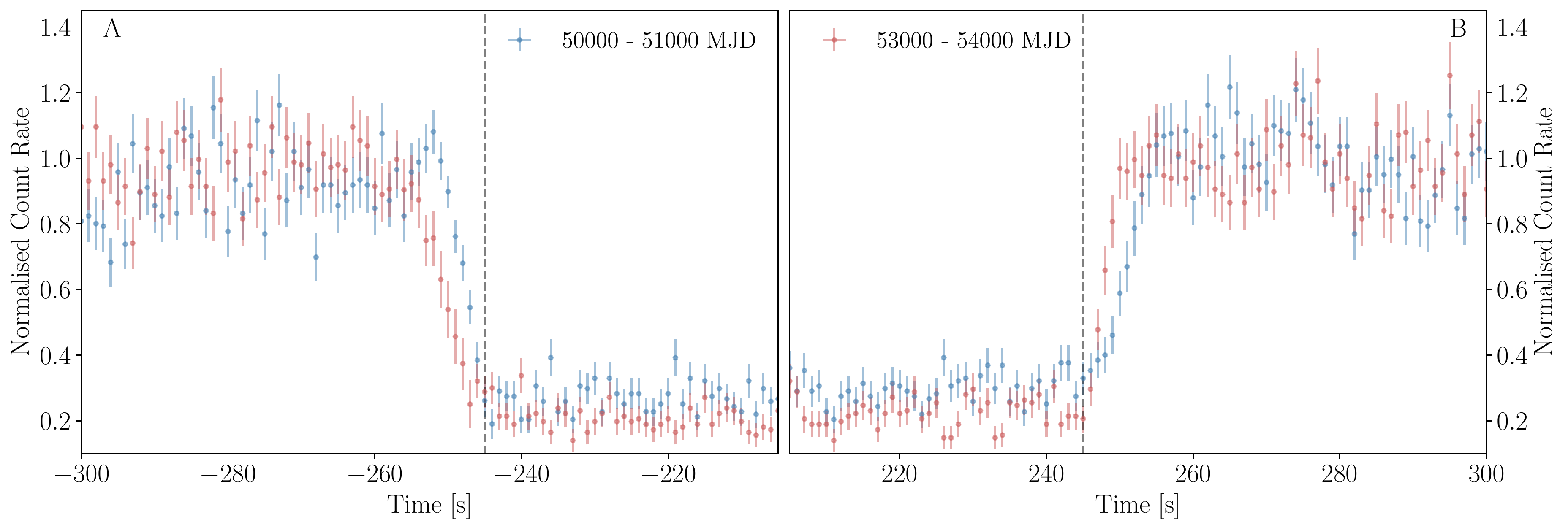}
\vspace{-0.5cm}
\caption{Stacked \textit{RXTE} eclipse profiles of EXO 0748 for two times periods, $50000 - 51000$ MJD (blue) and $53000 - 54000$ MJD (red). The eclipse profiles are normalised to have a mean out-of-eclipse count rate of $\sim 1.0$ and shifted horizontally such that the eclipses profiles in panel A (showing the ingress) are aligned at the $t_2$ times, and $t_3$ in panel B (showing the egress). In panels A and B respectively, the times $t_2$ and $t_3$ are represented by the black dashed lines. Clearly shown are the differences in ingress and egress duration between the two time periods.}
\label{fig:invseg}
\end{figure*}

After the aforementioned equilibrium state is reached, the eclipse transition durations and how they vary can provide insights into the presence of the absorbing medium around the companion, its radial extent and its structure. Utilising all available archival \textit{RXTE} observations of EXO 0748 that contain full eclipse profiles, we measure the duration of the ingress, egress and totality, thus uncovering how the transitions evolve over time. As there are 429 full \textit{RXTE} eclipses, we chose to fit the observed eclipse profiles with a simple eclipse profile model rather than our previously published eclipse model that describes the absorption by the ablated material layer (see \citealt{Knight2022a} and \citealt{Knight2022b}). As \textit{RXTE} does not extend to softer X-rays ($\lesssim 2$ keV), a simple eclipse model is sufficient as the majority of the absorption by the ablated layer occurs between $0.5 - 2.0$ keV.

Using the re-binned and normalised $2-15$ keV \textit{RXTE} eclipse profiles, we first identify the times $t_{90}$ and $t_{10}$ for both the ingress and egress.
The durations of the ingress and egress are initially determined as $\Delta t_{\rm{in}} \approx t_{10, \rm{in}} - t_{90, \rm{in}}$ and $\Delta t_{\rm{eg}} \approx t_{90, \rm{eg}} - t_{10, \rm{eg}}$, and the initial duration of totality is determined as $t_{\rm{tot}} \approx t_{10, \rm{eg}} - t_{10, \rm{in}}$. These measurements serve as both initial parameters for our eclipse fits and checks to ensure the eclipses have been correctly identified in each case, particularly for the observations in which the eclipse was not deliberately targeted. 

The four eclipse `contacts', $t_{1}$, $t_{2}$, $t_{3}$ and $t_{4}$ which are, respectively, the start of the ingress, start of totality, end of totality and end of the egress (e.g \citealt{Wolff2009}) are subsequently determined by fitting a simple `broken-line' eclipse model consisting of 5 parameters to the light curve. The model parameters are the mean pre-ingress count rate, $R_{\rm{in}}$, the mean post-egress count rate, $R_{\rm{eg}}$, the mean in-eclipse count rate, $R_{\rm{tot}}$, the gradient of the ingress $m _{\rm{in}}$ and the gradient of the egress $m_{\rm{eg}}$. Since we are working with normalised eclipse profiles, the first three parameters take values between $0.0 - 1.0$ and are used to fit the pre-ingress, post-egress and totality segments of the eclipse profiles. The gradients of the ingress and egress are determined by a linear fit through the predetermined points t$_{90}$ and $t_{10}$, which are fixed during the fit to the light curve. Each point of intersection between the five fitted lines thus corresponds to one of the four eclipse contacts. These contacts are related as $t_{2} = t_{1} + \Delta t_{\rm{in}}$, $t_{4} = t_{3} + \Delta t_{\rm{eg}}$ (e.g. \citealt{Wolff2009}), where the initial and minimum values of $\Delta t_{\rm{in}}$ and $\Delta t_{\rm{eg}}$ are set using the measurements of $t_{90}$ and $t_{10}$, as described above. Similarly, the totality duration is now determined as $\Delta t_{\rm{tot}} = t_{3} - t_{2}$ after starting from initial and maximum approximation above and the mid-eclipse time is $t_{\rm{mid}} = (t_{3} - t_{2})/2$. An example of a resulting fit is provided in Fig. \ref{fig:ECFit}. The results of this fitting procedure are presented in Table \ref{tb:RXTE_Table} along with a corresponding reduced $\chi^{2}$ and null-hypothesis probability for each eclipse fit.

The measured ingress, egress and totality durations are displayed as a function of mid-eclipse MJD
in Fig. \ref{fig:RXTE_Times}, panels A, B and C respectively. The colours represent hardness ratio, included to indicate how spectral changes to the system may be influencing in/egress durations and thus, the ablation. Additionally, a 45-point moving average of these durations as a function of mid-eclipse MJD are provided in Fig. \ref{fig:RXTE_Av_Times}, panels A, B and C respectively, allowing the trends to be seen more clearly. The colours now represent root mean squared (RMS) variability amplitude, which is known to be a good indicator of spectral state \citep{Belloni2010,Heil2015}.

The ingress and egress durations for EXO 0748 are are found to vary dramatically during the \textit{RXTE} era. The most striking characteristics can be seen in Fig. \ref{fig:RXTE_Av_Times}. The variations can inform on the quantity and density of ablated material near to the companion star; longer transitions indicate gradual X-ray absorption through lower density and/or extended material, while shorter durations indicate steep X-ray absorption through denser and/or less extended material. The ingress duration appears to roughly correlate with fractional RMS (with higher RMS corresponding to a harder spectral state) and anti-correlates with the egress duration. For most of the \textit{RXTE} era, the egress is longer than the ingress, except for MJD $\sim$53000-54000 during which this relationship flips. We illustrate this reversal in Fig. \ref{fig:invseg} by showing the stacked eclipse profiles from two time periods during the \textit{RXTE} era; MJD $50000 - 51000 $ (blue) during which the ingress is shorter than the egress and MJD $53000 - 54000$ (red) during which the ingress is longer than the egress. This reversal of eclipse asymmetry may be caused by the ablated material moving within the gravitationally bound region of the binary and/or by a change in the spectral state of the irradiating source that is driving the outflow from the stellar surface. For example, ablated material trailing behind the companion star due to the binary's orbital motion may `catch up' with the front of the companion star and extend the ingress duration. Also, a hard spectrum with a drop in the soft X-ray emission should be accompanied by a reduction in the outflow mass loss rate from the companion star. Such a reduction could explain why the egress duration decreases when the system transitions to a harder spectral state (higher RMS). Thus, the eclipse asymmetry reversing is a strong indication that the transition durations are being governed by irradiation driven, super-Roche lobe material.

In eclipsing redback and black widow pulsars the egress is typically longer than the ingress due to the orbital motion of the binary inducing a material trail on the egress side of the star. However, reversal of the eclipse asymmetry has previously been observed. One such case is the black widow pulsar PRS J2051$-$0827, where \citet{Polzin2019a} uncovered changes in the eclipsing material that occurred on $\sim$ month timescales. \citet{Polzin2019a} studied the evolution of the dispersion measure (DM) structure in orbital phases that correspond to the eclipses, during a decade long study. They find the DM peaks on the ingress side between 2011 and 2014, but later shifts to the egress side, thus implying that the eclipsing material's structure can change on timescales similar to those observed for EXO 0748.

\vspace*{-0.4cm}
\subsection{Totality and Orbital Period Variations}
\begin{figure}
\centering
\includegraphics[width=\columnwidth]{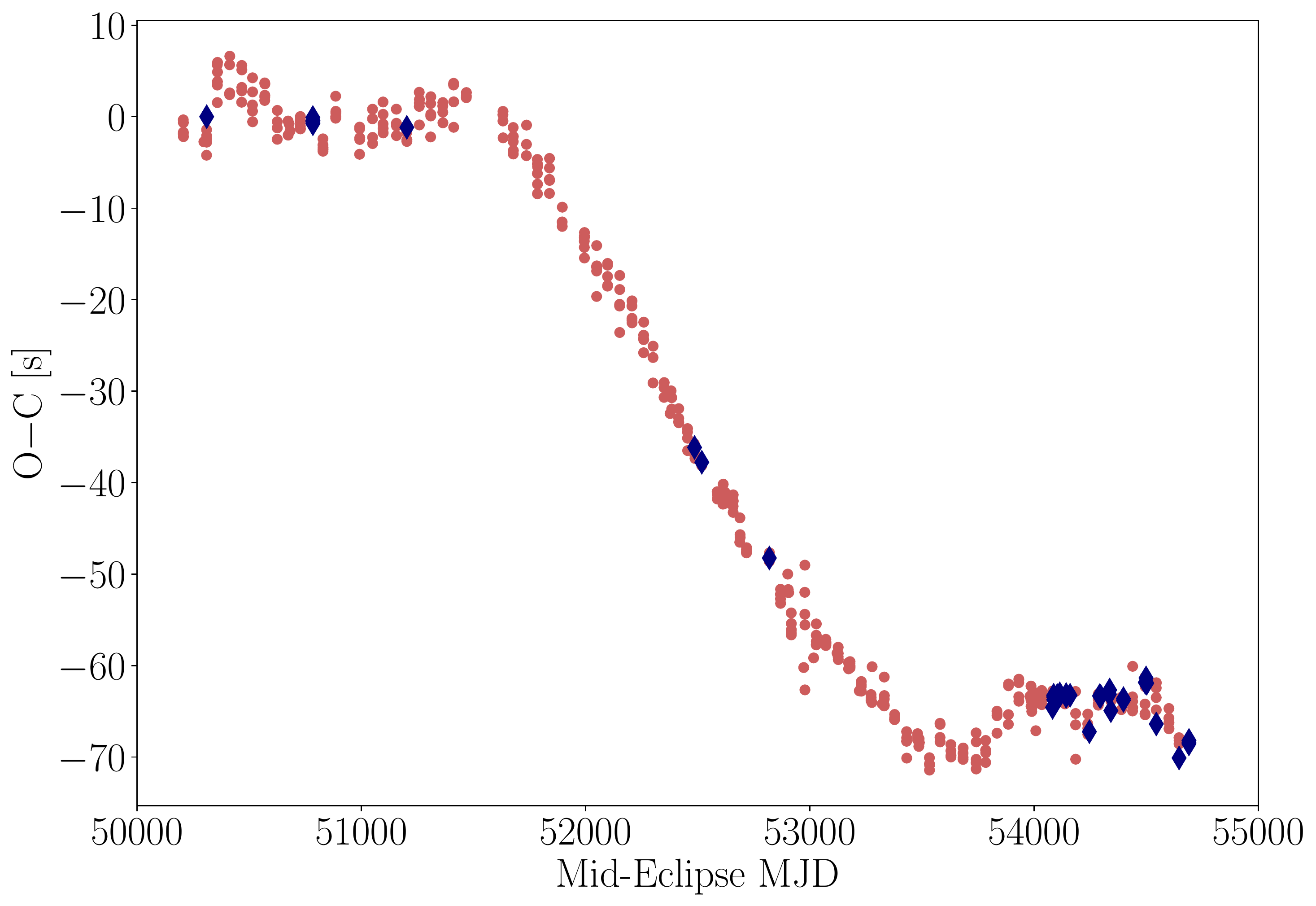}
\vspace{-0.5cm}
\caption{Measured O$-$C residuals of the mid-eclipse times for each of the full eclipses of EXO 0748 observed by \textit{RXTE}. These residuals confirm that EXO 0748 showed three constant orbital period solutions during the \textit{RXTE} era, originally uncovered by \citet{Wolff2009}. Red dots correspond to ObsIDs originally considered by \citet{Wolff2009} and blue diamonds represent additional ObsIDs considered in this work.}
\label{fig:O-C}
\end{figure}

\begin{figure}
\centering
\includegraphics[width=\columnwidth]{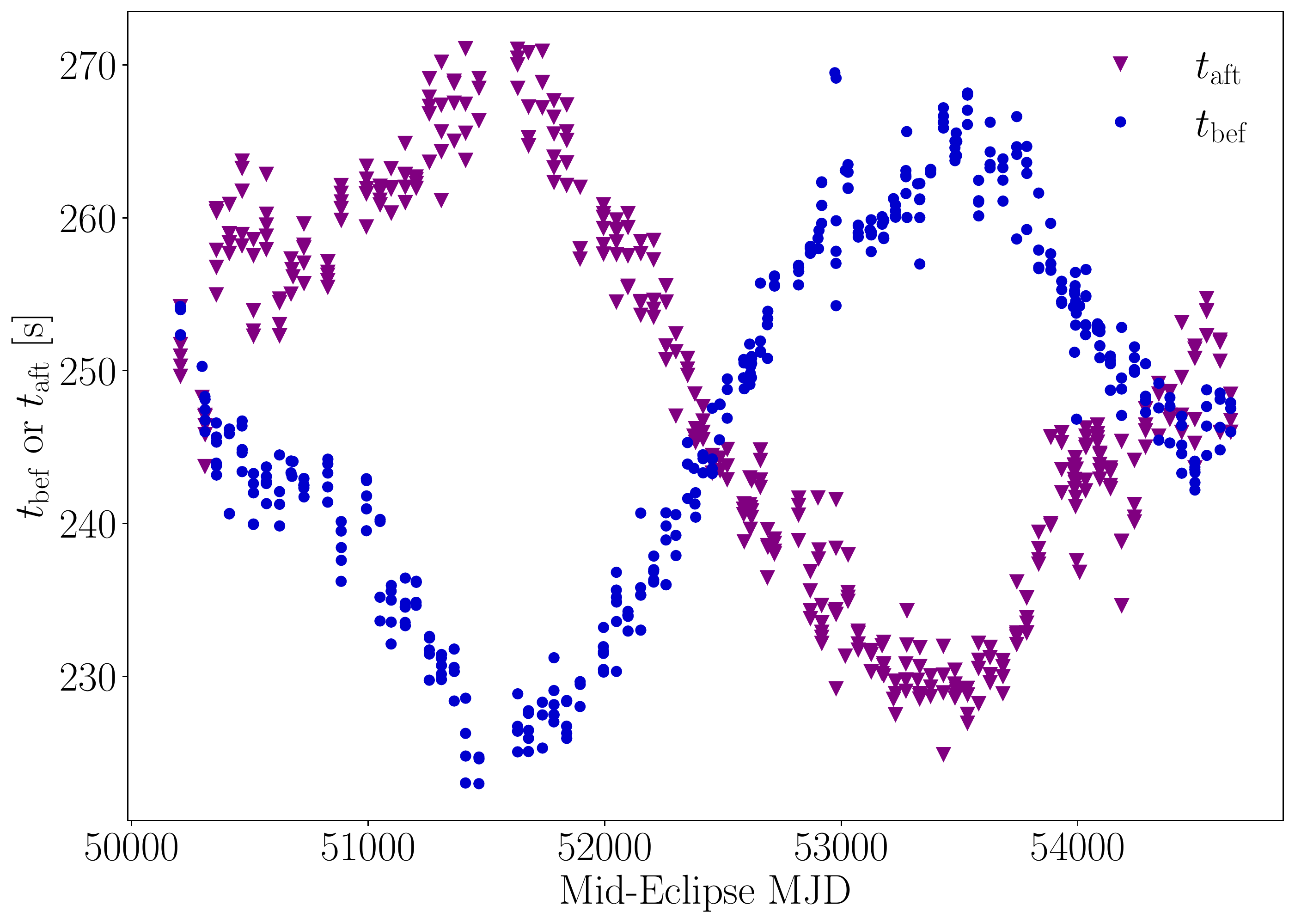}
\vspace{-0.5cm}
\caption{Calculated values of $t_{\rm{bef}} = (t_{e} / 2) - (\rm{O-C})$ and $t_{\rm{aft}} = (t_{e} / 2) + (\rm{O-C})$ as a function of mid-eclipse MJD for all full \textit{RXTE} eclipses of EXO 0748. Here, the orbital period is defined such that the O$-$C residuals at the start and end of the \textit{RXTE} era are zero, and the period is constant. As the difference between the maximum and minimum values of both $t_{\rm{bef}}$ and $t_{\rm{aft}}$ are $\sim 45$ s, the ablated material cannot explain the observed O$-$C residuals.
}
\label{fig:t_half}
\end{figure}

The evolution of the totality duration for EXO 0748 during the \textit{RXTE} era is shown in Fig. \ref{fig:RXTE_Times} C, covering a range $\sim 485 - 510$ s, and the corresponding 45-point moving average in Fig. \ref{fig:RXTE_Av_Times} C. These measurements are generally consistent with those presented by \citet{Wolff2009}, who also studied the variations in the duration of totality (see Fig. 5 of \citealt{Wolff2009}). The \textit{RXTE} era initially shows a decrease in the totality duration and continues to drop until $\sim$ 52500 MJD. The system is in an a softer state during this period. Following this, the totality duration fluctuates but increases on average. These fluctuations are accompanied by a varying, but overall harder spectral state.

We first investigate if this evolution in totality duration can be explained by evolution of the binary system. If we assume that the companion star is filling its Roche lobe, with no super-Roche lobe material, the duration of totality, $t_e$, relates to the orbital period, $P$, mass ratio, $q=M_\textrm{cs}/M_\textrm{ns}$ and binary inclination, $i$ via the formula
\begin{equation}
    \sin i = \frac{ \sqrt{ 1 - h^2(q) } } { \cos( \pi t_e / P ) },
    \label{eqn:sini}
\end{equation}
where $h(q)$ is the ratio of the Roche-Lobe radius to the orbital separation \citep{Knight2022a, Knight2022b}.
The inclination is unlikely to display the drastic changes needed to explain the observed changes in the totality duration and the orbital period is constant in epochs during which the totality duration varies \citep{Wolff2009}. Therefore, in this picture only the mass ratio can drive the observed changes in totality duration.

The orbital period of EXO 0748 is $3.824$ hrs \citep{Parmar1991, Wolff2009} and the binary is inclined to $\sim 77 ^\circ$ \citep{Knight2022a}. For the observed range of totality durations of $\sim 485 - 510$ s, the corresponding range in mass ratio is $q \sim 0.215 - 0.224$ (\citealt{Knight2022a} measured $q \sim 0.222$). Assuming that the change in mass ratio is driven by accretion onto the NS (i.e. mass transfer $>$ mass loss), then the initial decrease in totality duration from $\sim 510$ s to $\sim 485$ s, which occurs over a $\sim 5$ year period, requires the companion to lose $\sim 3.3\%$ of its mass to the NS, which then grows by $\sim 0.7\%$. If $M_\textrm{ns} \approx 2~M_\odot$ \citep{Knight2022a}, this corresponds to a mass transfer rate of $\sim 3\times 10^{-3} ~M_\odot/\textrm{yr}$. The rest mass energy transferred is therefore $\dot{M} c^2 \approx 1.7 \times 10^{44}~\textrm{erg}$. 
For a reasonable radiative efficiency, this predicts a luminosity far in excess of what is observed. Therefore, we can rule out the observed changes in totality duration being entirely driven by changes in the Roche potential. Moreover, it would be very difficult to explain $t_e$ \textit{increasing} again because, in this picture, we would require mass to be transferred from the NS to the companion.

An alternative explanation for the variations in totality duration is that a portion of totality is sometimes a result of optically thick material beyond the Roche Lobe of the companion. Specifically, our sight line through the ablated material becomes optically thick near the companion star radius, cutting out all light and extending totality. In this interpretation, the time it actually takes the companion's Roche Lobe to cross our line of sight is $\lesssim 491$s, with the ablated material adding as much as $\sim 9$ s to totality in the case of the longest observed eclipses early in the \textit{RXTE} observations. A true totality duration of $\sim 491$s corresponds to a $\sim 2 \%$ reduction of $t_{e}$ assuming an average totality duration of $\sim 500$s. Following equations 9 and 17 from \citet{Knight2022a}, we see that $M_{\rm{ns}} \propto (\cos^{3}(\pi {t_{e}/P}))^{-1}$, and $t_e$ does not influence the K-correction (the deviation between the reprocessed light centre and the centre of mass of a Roche lobe-filling star; \citealt{MunozDarias2005}). Therefore, a $2 \%$ decrease of $t_{e}$ corresponds to a $5 \%$ increase in $M_{\rm{ns}}$ which is a negligible change to the NS mass (\citealt{Knight2022a} present a $\sim 14 \%$ uncertainty on their NS mass of $ 2 M_{\odot}$), so this could explain the variations in totality duration. However, totality does not need to correspond to the surface of the Roche lobe. For example, totality could correspond to an interval shorter than the size of the Roche lobe if enough mass can get across the inner Lagrange point to produce the outbursts while the same material layer seen in the eclipses appears optically thin to X-rays. Calculating this, however, is beyond the scope of this work.

The orbital period of EXO 0748 displayed several sharp transitions between different, seemingly constant values over the course of its outburst \citep{Wolff2009}. Fig. \ref{fig:O-C} shows the O$-$C residuals\footnote{Observed centre of totality minus that predicted for a constant orbital period.} as a function of mid-eclipse MJD for the \textit{RXTE} eclipses of EXO 0748. We use our measurements of the mid-eclipse MJD (see Table \ref{tb:RXTE_Table}) to recreate Figure 3 from \citet{Wolff2009}, confirming that during the \textit{RXTE} era there are three separate, constant orbital period solutions. We see that the new O$-$C measurements are consistent with the previously uncovered relationship. We investigate whether this behaviour can be explained by assuming a constant orbital period with the ablated material causing the totality duration to appear longer or shorter. We define $t_{e} = t_{\rm{bef}} + t_{\rm{aft}}$ as the duration of totality, where $t_{\rm{bef}} = (t_{e} / 2) - (\rm{O-C})$ and $t_{\rm{aft}} = (t_{e} / 2) + (\rm{O-C})$. Therefore, by calculating $t_{\rm{bef}}$ and $t_{\rm{aft}}$, we can see if $t_{\rm{mid}}$, the time at the centre of the eclipse, is being shifted by the start or end of totality arriving earlier or later than expected due to excess, optically thick material close to the star’s surface. We find that the ablated outflow would need to be responsible for $\sim 45$ s of the totality duration in order for this hypothesis to be true (see Fig. \ref{fig:t_half}), meaning that the true duration of totality would be $\sim 455$ s. This amount of time is too large to be accounted for by the ablated material layer, especially as the ablated outflow is already responsible for extending the eclipse transitions (which are $\sim 15-20$ s long \citealt{Knight2022a}). This hypothesis also predicts that the O-C residuals correlate with the in/egress durations, however, we do not find any evidence for this.
We therefore conclude that the observed O$-$C residuals really are driven by changes in orbital period, which we speculate could be caused by a circumbinary disc and mass loss from the second Lagrange point  (see e.g. \citealt{Heath2020,Avakyan2021}).

\begin{figure}
\centering
\includegraphics[width=\columnwidth]{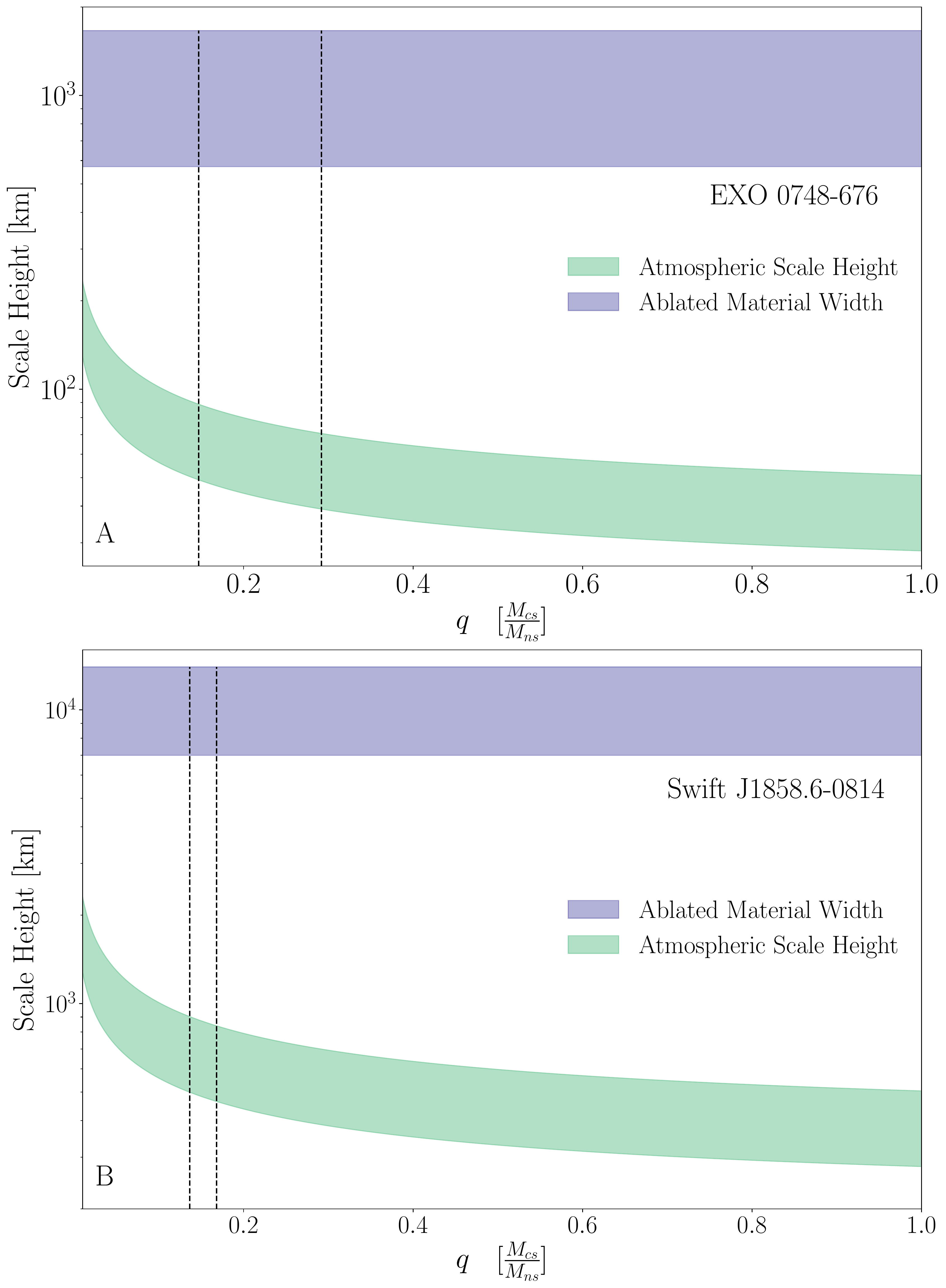}
\vspace{-0.5cm}
\caption{The width of the material layer around the companion star measured via eclipse mapping for EXO 0748 \citep{Knight2022a} (Panel A, blue) and Sw J1858 \citep{Knight2022b} (Panel B, blue) compared to the width of the companion star atmosphere calculated using Equation \ref{eq:hsc}, which assumes the star is in hydrostatic equilibrium (atmosphere is supported by gas pressure) for EXO 0748 (Panel A, green) and Sw J1858 (Panel B, green) respectively. In both sources we conclude the the atmospheric scale height is not sufficient to explain the radial extent of the obscuring material around the companions.
In each panel, black dashed lines indicate the mass ratio in each source measured via eclipse mapping \citep{Knight2022a,Knight2022b}.
}
\label{fig:Scale_Height}
\end{figure}

\section{Discussion}
\label{sx:Discuss}
We have investigated two eclipsing NS LMXBs -- EXO 0748 and Sw J1858 -- that both show evidence that their companion stars are undergoing ablation by X-ray irradiation from the NS. The physical scenario of ablation was initially suggested by \citet{Parmar1991} to explain the highly variable and extended eclipse transition durations in some \textit{EXOSAT} observations of EXO 0748. The same scenario was later used to explain the eclipse profiles of both EXO 0748 and Sw J1858, as their extended ingress and egress profiles are well described by a model that includes optically thin material around the companion star \citep{Knight2022a, Knight2022b}. The eclipses of both EXO 0748 and Sw J1858 are asymmetric, respectively, with ingress and egress durations of $\sim 15$s and $\sim 17$s and $\sim 106$s and $\sim 174$s, which are reproduced well by modelling the optically thin material around the companion star with an exponential or Gaussian radial density profile \citep{Knight2022a, Knight2022b}. We can make a simple calculation of the atmospheric scale height of each of the companions stars assuming they have isothermal hydrostatic atmospheres supported by gas pressure. In this case, the density decreases exponentially with height, $\rho\propto\exp(-h/h_{\rm sc})$, as inferred by \citet{Knight2022a} and \citet{Knight2022b}. Thus, the scale height of an isothermal companion star atmosphere, in the simplest case, is given by
\begin{equation}
    h_{\rm sc} = \frac{R_{\rm cs}^2kT}{GM_{\rm cs}\mu m_{\rm p}}.
    \label{eq:hsc}
\end{equation}
Here $R_{\rm cs}$ is the Roche lobe radius, $k$ is the Boltzmann constant, $T$ is the temperature of the atmosphere, $G$ is the gravitational constant, $m_{\rm{p}}$ is the mass of a proton and $\mu$ is the mean particle mass in AMU. For $3000 \rm{K} \leq T \leq 7000 \rm{K}$, $\mu = 1.0$ and $1.4 M_{\odot} \leq M_{\rm{NS}} \leq 3.0 M_{\odot}$, we find that the atmospheric scale heights of both companion stars are far less than the radial extent of the ablated material layers inferred via eclipse mapping (see Figs. \ref{fig:Scale_Height} A and B respectively for EXO 0748 and Sw J1858). Therefore, this simple calculation strongly supports the presence of super-Roche lobe material in both systems and implies that the atmospheric scale height of the companion is negligible when discussing the X-ray absorbing layer. The atmospheric (gas pressure) scale height could still be an important factor if X-ray heating, which is a component of ablation, significantly heats the atmosphere.

Here, we have conducted orbital phase-resolved spectroscopy of both sources to trace the ablated material out to much larger distances from the companion star. For both objects, we find that the ionized absorbing material is still present beyond the eclipses, at orbital phases $|\phi| \gtrsim 0.2$ rad). We find that the column density, ionization and covering fraction drop off with distance from the companion star. These findings inform us that the ablated material remains present at large distances from the companion star and is distinguishable from the ISM. In addition, the measured ionization is consistent with the outflow being driven by irradiation and the evolution of the covering fraction suggests that the ablated material becomes clumpy when diffusing away from the companion star. This inferred geometry is illustrated in Fig. \ref{fig:Ablation_Schematic}.

We have further investigated the long-term evolution of eclipses in both objects. Since there are hundreds of archival eclipses for EXO 0748 and only $12$ partial eclipses for Sw J1858, our analysis focuses mainly on the former object. Utilising all available \textit{RXTE} observations of EXO 0748, we find several properties that cannot be explained without the presence of stellar material extending beyond the companion's Roche Lobe. First, the evolution of the totality duration, which cannot be driven by changes to the binary system, but must instead be due to ablated material close to the companion star surface sometimes becoming sufficiently optically thick to extend totality. Second, the evolution of the ingress and egress durations; for most of the time \textit{RXTE} was monitoring EXO 0748's eclipses, the egress was longer than the ingress, indicating that material was being dragged behind the companion star on its orbit. However, for several years this behaviour reversed, with the ingress becoming longer than the egress, similar to the observed behaviour of the known black widow pulsar PRS J2051$-$0827 \citep{Polzin2019a}. We further find that the ingress and egress durations appear to be anti-correlated with one another and we uncover tentative evidence that the transition durations are correlated with the RMS variability amplitude of the irradiating X-ray signal. This is well known as a good proxy for spectral state \citep{Belloni2010, MD2010, Heil2015}. We, therefore infer, that the ingress is longer when a harder spectrum is being emitted from the NS vicinity, while longer egresses align with irradiation by a softer spectrum, thus providing further evidence that X-ray irradiation is responsible for the presence of the ionized absorbing material in the vicinity of the companion.

While EXO 0748 and Sw J1858 display remarkably similar eclipse and orbital phase-resolved spectral properties, there are also notable differences between the two systems. First, eclipse mapping analyses indicates that the companion star in EXO 0748 is consistent with being a main sequence star (mass: $\sim 0.4 M_{\odot}$, radius: $\sim 0.43 R_{\odot}$; \citealt{Knight2022a}) but implies that the companion star in Sw J1858 is far less compact than a typical main sequence star (mass: $0.183 M_{\odot} \leq M_{\rm{cs}} \leq 0.372 M_{\odot}$, radius:$1.02 R_{\odot} \leq R_{\rm{cs}} \leq 1.29 R_{\odot}$; \citealt{Knight2022b}). Therefore if ablation has influenced the global properties of the companion star in these two objects, it has not caused the companion in EXO 0748 to diverge far from the expected main sequence mass-radius relation. 

Secondly, Sw J1858 has a longer orbital period than EXO 0748 ($\sim 21.3$ hrs versus $3.82$ hrs). It is intuitive to associate ablation with more compact systems e.g. short-period binaries since the irradiating X-ray flux is higher for a given source luminosity, so ablation may be more effective in EXO 0748 than in Sw J1858. \citet{Podsiadlowski1991a} showed that a low-mass companion star in an accreting binary could be induced to expand by a factor of 2 to 4 and lose mass by an irradiating flux of $\log(F_x / \rm{erg~s}^{-1} \rm{cm}^{-2}) \approx 11.6$. For reasonable distance estimates to each source ($6.8$ kpc for EXO 0748 \citealt{DiazTrigo2011} and $12.8$ kpc for Sw J1858 \citealt{Buisson2020} ), the irradiating flux can be estimated to be above this threshold for both EXO 0748 ($\log(F_{x} / \rm{erg~s}^{-1}~\rm{cm}^{-2}) \sim 12.8$) and Sw J1858 ($\log(F_{x} / \rm{erg~s}^{-1} \rm{cm}^{-2}) \sim 12.1$; \citealt{Knight2022b}). Therefore, although the orbital period is very different between the two systems, both appear to have enough irradiating flux to drive expansion/bloating of the companion's atmosphere and contribute to mass loss. We therefore suggest that other LMXBs with an irradiating flux $\gtrsim 10^{12}~\rm{erg~s}^{-1}~\rm{cm}^{-2}$ could host material ablated from the companion star outside of the Roche-equipotential boundary. Note, however, that these simple calculations assume that the emission is isotropic and have not been averaged over an outburst duty-cycle. As such, these values represent upper limits on the irradiating flux incident on the companion star. Proper consideration of the incident flux requires us to compute the average flux over a timescale relevant to the star's response. However, this is challenging as both EXO 0748 and Sw J1858 have only showed one outburst each, so the duration of each source's outburst cycle is unknown. Reasonable order-of-magnitude estimates for the duty cycles of X-ray transients fall within the range $0.01 - 0.1$ \citep[e.g.][]{Barnard2014}, so the the true irradiating flux required for ablation is likely $< 10^{12}~\rm{erg~s}^{-1}~\rm{cm}^{-2}$. Further observations of ablation in X-ray binaries are required to constrain this flux boundary.

If we assume that the companion star has expanded due to X-ray irradiation, we would also need to consider whether the binary itself has expanded as a result. Such expansion could cause mass transfer to cease (e.g. a transient system) and thus, stop any ongoing irradiation-driven ablation. This scenario may explain the short outburst in Sw J1858 and inferred radius of its companion star, but is overall incompatible with EXO 0748. Furthermore, a bloated/expanded atmosphere remains challenging to reconcile with the large observed eclipse asymmetry. While stellar atmospheres can be asymmetric, this has only been observed in asymptotic giant branch (AGB) stars or pulsating red giants; e.g. Mira A \citep{Vlemmings2019}. Even then, the degree of asymmetry seen for these evolved stars is not sufficient to explain the $t_{\rm{in}} / t_{\rm{eg}} \sim 0.6$ asymmetry we observe in Sw J1858 and is even difficult to reconcile with the $t_{\rm{in}} / t_{\rm{eg}} \sim 0.87$ asymmetry observed in EXO 0748.

The asymmetric, photon energy-dependent eclipse profiles (compare Fig. 2 of \citealt{Knight2022a} and Fig. 2 of \citealt{Knight2022b}), the low-mass companion stars and the presence of ionized, optically thin material ablated from the companion stars in both sources are reminiscent of redback and black widow pulsars, in which the NS with a pulsar wind drives the ablation of the companion star, having previously been spun up by accretion \citep{Fruchter1988, Alpar1982, Rad1982}. While the number of known eclipsing redback and black widow pulsars remains low, the observed radio eclipses (see e.g. \citealt{Polzin2018} for discussion of the eclipses observed for J1810+1744) are similar in many ways to the X-ray eclipses we observe for EXO 0748 and Sw J1858. Additionally, the reversal in eclipse asymmetry that we have uncovered in the \textit{RXTE} eclipses of EXO 0748 is also seen in the known black widow pulsar PRS J2051$-$0827 \citep{Polzin2019a}. For PRS J2051$-$0827, the structure of the ablated material changes on $\sim$ month timescales, peaking on the ingress side for $\sim$ 3 years, before reversing and peaking on the egress side. These timescales are comparable to those uncovered for EXO 0748 thus increasing our confidence that super-Roche lobe material is influencing the observed eclipse profiles in both systems. Furthermore, the presence of very broad optical emission lines \citep{Ratti2012} and a broad C IV (UV) feature \citep{Parikh2020} observed from EXO 0748 while in quiescence, provide additional evidence that EXO 0748 is spider-like. \citet{Ratti2012} attributed the broad optical emission lines to an outflow driven by X-ray irradiation and/or a pulsar wind, while \citet{Parikh2020} found similarities between the broad C IV (UV) feature and observed features of the known tMSP PSR J1023$+$0038.

These findings contradict analysis by \citet{Church2004}, who interpret the eclipse ingress and egress of EXO 0748 as being due to the transit by the companion star across a very large X-ray emitting region. Such an interpretation is incompatible with the observed asymmetry and photon energy dependence of the eclipses. Specifically, the eclipse of an extended X-ray emitting region cannot explain how the soft X-ray ingress is observed to start earlier and the egress to finish later than the hard X-ray ingress and egress, while the start and end of totality are both independent of photon energy (see Section 2.3.1 of \citealt{Knight2022a}). In contrast, the eclipse profiles and phase-resolved spectra are naturally explained by the presence of ionizing material ablated from the companion star.

Despite there being no additional observed similarities between Sw J1858 and black widow pulsars, its evolved companion is consistent with observations of known huntsman pulsars \citep{Strader2015} suggesting Sw J1858 is huntsman-like. However, its short orbital period is inconsistent with huntsman systems and is instead, similar to redback systems. Sw J1858 also shares a number of properties with EXO 0748, including but not limited to the X-ray eclipse characteristics, out-of-eclipse flaring and dipping behaviour \citep{Bonnet-Bidaud2001, Homan2003, Buisson2021}, Type I X-ray bursts \citep{Gottwald1986, Bonnet-Bidaud2001, Buisson2020} and a stellar prominence during some ingresses \citep{Wolff2007, Buisson2021}. We consider these similarities sufficient to group the two systems at this stage. Interestingly, \citet{Church2004} identify several other eclipsing NS XRBs with even longer transition durations than those seen in EXO 0748 (see Fig. 2 of \citealt{Church2004}), implying that ablation is common in binaries that are compact enough for eclipses to be likely and that it becomes a stronger effect with increasing X-ray luminosity.

The aforementioned similarities between these X-ray observations and the radio observations of eclipsing spider pulsars led us to term EXO 0748 and Sw J1858 as \textit{false widows}. We define a false widow as a LMXB whose companion star is being ablated by X-ray irradiation from the NS and accretion disc and/or a pulsar wind. For EXO 0748, some observations indicate the presence of a pulsar wind (see \citealt{DiazTrigo2011, Parikh2020}), but the response of the transition durations to spectral state strongly suggest that the outflow is irradiation-driven. We further suggest that an irradiating flux on the companion from the primary $\gtrsim 10^{12}~\rm{erg~s}^{-1}~\rm{cm}^{-2}$ is sufficient to drive ablation in such systems. Uncovering ablation in other LMXBs will enable us to test this boundary and assess the commonality of ablation in XRBs.

We speculate that false widows may represent progenitors of redback pulsars under the assumption that ablation begins while the source is actively accreting and continues into the rotation powered phase. At the time of writing, the false widows are not known to display pulsations at any wavelength, however the distance to each source ($\approx 13$ kpc for Sw J1858 \citep{Buisson2020} and $\approx 7$ kpc for EXO 0748 \citep{DiazTrigo2011}) makes detecting pulsations challenging. The spin of EXO 0748 is known to be $552$ Hz \citep{Galloway2010}, which would place it within the millisecond regime, like redbacks and black widows, if it is a pulsar, but the spin of the NS in Sw J1858 is not known at the time of writing. While both sources are in X-ray quiescence we will search for radio pulsations. Additionally spider pulsars are known to harbour some of the most massive NSs \citep{Romani2012, Linares2018, Burdge2022} which is consistent  with the $\sim 2 M_{\odot}$ NS in EXO 0748 \citep{Knight2022a}. Our speculations could be settled by understanding whether ablation occurs solely in NS XRBs or whether it also occurs in BH XRBs. Evidence for the latter would not only imply that irradiation is driving ablation but also suggest that ablation is ubiquitous among XRBs, thus contradicting our hypothesised link between false widows and redback pulsars. The presence of ablated material is of course harder to infer in the absence of eclipses, but may still be possible to detect with orbital phase-resolved spectroscopy even for non-eclipsing sources and observed changes in $N_{\rm{H}}(t)$ could constrain the orbital period.

\section{Conclusions}
\label{sx:Conclude}
The X-ray eclipses exhibited by two NS LMXBs, EXO 0748$-$676 and \textit{Swift} J1858.6$-$0814, share several characteristics such as asymmetry and photon energy dependence throughout the eclipse transitions. Additionally, the two sources share some out-of-eclipse behaviours such as type I X-ray bursts, dips and flares. We utilised archival data to further explore the similarities between the two systems through phase-resolved spectroscopy of their near-eclipse epochs. We determined that an ionized absorber exists close to the companion star in both systems. The column density and covering fraction of the ionized absorber decreases further from the companion star, implying that the material becomes clumpy at large distances. This is consistent with the optically thin material layer inferred to extend the duration of the eclipse transitions of both LMXBs \citep{Knight2022a, Knight2022b}. This clumpy material arises from the ablation of the companion star's atmosphere by X-ray irradiation and/or a pulsar wind, similar to the cannibalistic behaviour exhibited by redback and black widow pulsars. We therefore refer to EXO 0748$-$676 and \textit{Swift} J1858.6$-$0814 as \textit{false widows}, defined as an XRB hosting an ablated companion star. We speculate that the false widows may represent progenitors of redback pulsars under the assumption that ablation begins during the accretion-powered phase and continues into the rotation-powered phase. This mechanism could provide a way to create the under-massive companions in short-period binaries like those observed within spider pulsars \citep{Fruchter1988, Stapper1996}. By modelling all available full eclipse profiles of EXO 0748$-$676 from the \textit{RXTE} data archive, we uncover a reversal of the eclipse asymmetry. This appears, tentatively, to depend on the spectral state of the system and could also be influenced by the movement of ablated material within the gravitationally bound region of the binary, similar to observations of the black widow pulsar PRS J2051$-$0827. We can evidence the link between the false widows and spider pulsars by searching for radio pulsations from the two sources while they are in X-ray quiescence and determining whether ablation is ubiquitous among XRBs.

\section*{Acknowledgements}
A. K. acknowledges support from the Oxford Hintze Centre for Astrophysical Surveys, which is funded through generous support from the Hintze Family Charitable Foundation. A. I. acknowledges support from the Royal Society. J.v.d.E. is supported by a Lee Hysan Junior Research Fellowship awarded by St. Hilda's College, Oxford. L.R. acknowledges support from STFC who has funded their research. This research has made use of software and data provided by the High Energy Astrophysics Science Archive Research Center (HEASARC) and the SAO/NASA Astrophysics Data System. The authors are grateful to the anonymous referee for a thorough and insightful report that prompted valuable changes to the manuscript.

\section*{Data Availability}
The data used in this study are publicly available from the HEASARC website. The models used are available upon reasonable request to the authors.



\bibliographystyle{mnras}
\bibliography{All_Refs} 



\begin{appendices} 
\section{RXTE Data Table for EXO 0748$-$676}
\label{sx:RXTE_Table}
The \textit{RXTE} data archive contains a total of 746 observations of EXO 0748$-$676, taken between March 1996 and September 2010, which contain 429 full eclipse profiles. The results of fitting the simple eclipse model described in Section \ref{sx:Transitions} to these data are provide in Table \ref{tb:RXTE_Table}, which details the Obs ID, number of active PCUs (proportional counter units), mid-eclipse MJD, ingress duration, egress duration, totality duration, reduced $\chi^{2}$, null-hypothesis probability, mean $2 - 15$ keV out-of-eclipse count rate and hardness ratio. Each of the durations are provided with a $1 \sigma$ confidence interval and the hardness ratio is calculated as ${F_{10-16 \rm{keV}}}/{F_{6-10 \rm{keV}}}$. 

\section{Phase Resolved Spectroscopy}
\label{sx:prs}
The principal result of this work, respectively shown in Figs. \ref{fig:EXO_PRS} and \ref{fig:J1858_PRS}, is the phase-resolved absorption spectroscopy of the near-eclipse epoch of the two false widows EXO 0748 and Sw J1858. From this analysis we uncover evidence for an ionised absorbing medium surrounding the companion star in both LMXBs which supports our hypothesis that in both EXO 0748 and Sw J1858, the companion stars are experiencing irradiation driven ablation. For each source, we have modelled 44 phase-resolved spectra within \textsc{xspec} using our local absorption model \textsc{abssca}, which introduces the hydrogen column density, $N_{\rm{H}}$, the log of the ionization, $\log(\xi)$ and covering fraction as properties of the companion star's outflow. Any absorption from the ISM is modelled separately using the \textsc{xspec} model \textsc{tbabs}, assuming the abundances of \citet{Wilms2000}. The three parameters introduced by \textsc{abssca} each create subtly different changes to the spectra. The material hydrogen column density, $N_{\rm{H}}$, alters the transmission factor such that higher values of $N_{\rm{H}}$ increase the material optical depth by an energy-independent factor. Since the absorption cross-section typically decreases with energy, increasing $N_{\rm{H}}$ typically has the effect of preferentially suppressing soft X-rays. The ionization parameter, $\log(\xi)$, impacts the number of bound atomic species in the absorber which in turn impacts the energy-dependence of the absorption cross-section. The covering fraction, f$_{\rm{cov}}$, changes the absorption fraction such that lower covering fractions allow more photons to pass through and thus increases the peak of the spectrum. Since each of these effects are only subtly different, we must determine whether all three parameters can be individually constrained, particularly if the spectra are not of high quality.

We begin by assessing the quality of each spectrum by checking the count rate for each bin of each spectrum. While we have opted for wide energy bands to increase the number of counts per bin (see Section \ref{sx:PRS}), in some cases the number of counts remains too low to perform the fits using chi-squared statistics. Indeed, when testing the chi-squared fits we often found very tight, unbelievable parameter constraints or that a comparable fit could be found with entirely different parameters. We instead opt to use C-statistics (hereafter c-stat) within \textsc{xspec} as it is more appropriate for fitting lower quality data with fewer counts \citep{Arnaud1996} and did not encounter the aforementioned issues. Note, that while binning the data to ensure a reasonable number of counts per bin is not required when fitting with c-stat, we do not re-bin the data into finer energy bands. This ensures consistency between the analysis presented here and our earlier phase-resolved spectral analyses of the same sources \citep{Knight2022a, Knight2022b}. For each source, we then select four spectra, two from the ingress or pre-ingress phases and two from the egress or post-egress phases to represent the spectral fits and investigate the partial parameter degeneracies.

\begin{figure*}
\centering
\includegraphics[width=\textwidth]{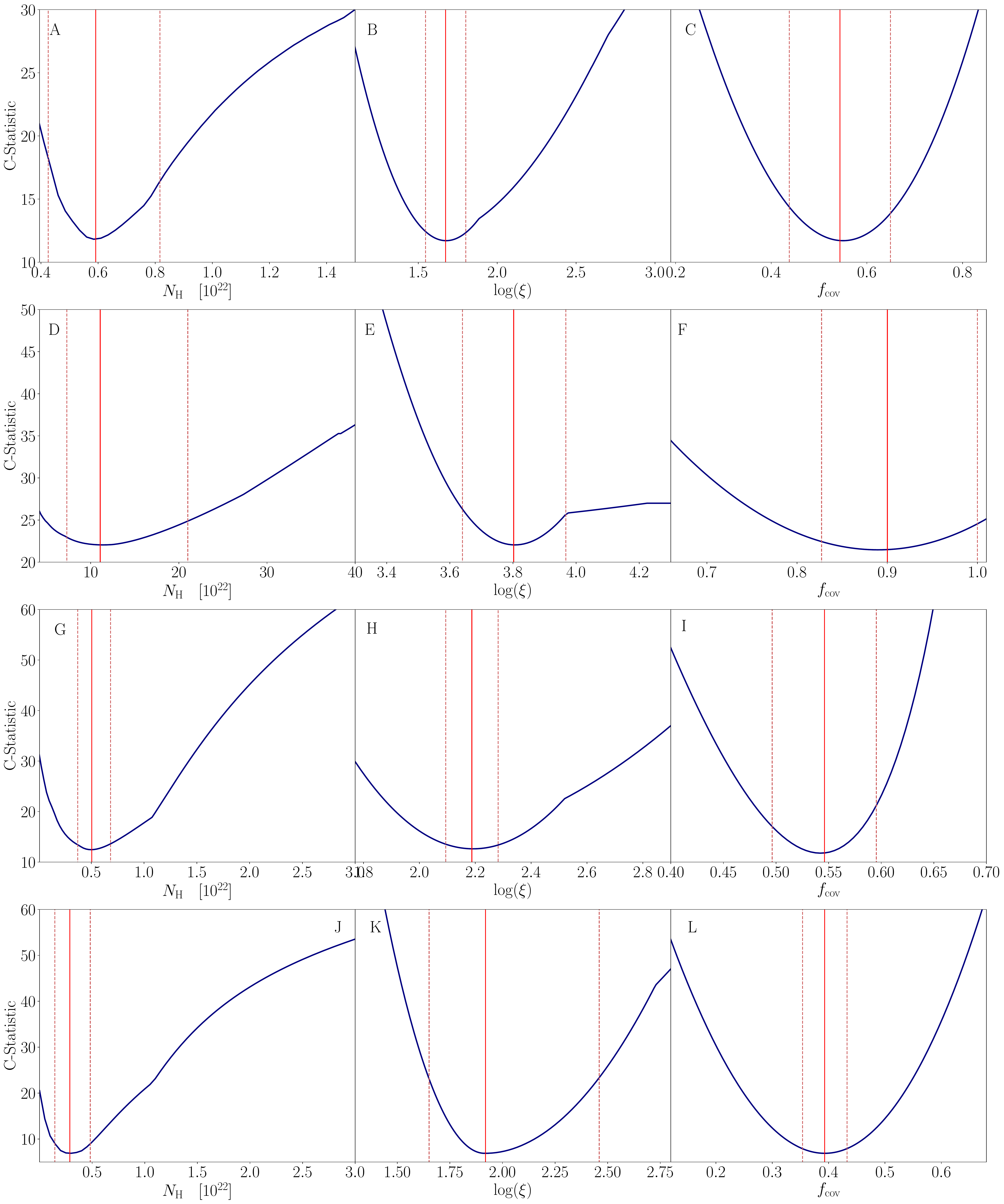}
\vspace{-0.5cm}
\caption{1D parameter contours (blue) for four different phase-resolved spectra of EXO 0748 obtained by running \texttt{steppar} analysis within \textsc{xspec}. Respectively, panels A-C, D-F, G-I and J-L correspond to Spectra 8, 18, 36 and 44, following a left to right numbering convention for the spectral fits in Fig. \ref{fig:EXO_PRS}. For each spectra $\log(\xi)$, f$_{\rm{cov}}$ were respectively varied in the ranges $1.0 - 3.0 $ and $0.05 - 0.95$ using 400 steps, except panel E, which instead used a range of $2.5 - 4.5$. For each spectra, $N_{\rm{H}}$ is varied through a range of values surrounding the best fitting values, again using 400 steps. For panels A, D and J this range is $1.0 \times 10^{-5} - 5.0$ while the range used in panel G is $0.1 - 50.0$. In each panel the best fitting parameter value is shown by the solid red line while the dashed red lines represent $1 \sigma$ contours. 
}
\label{fig:ex10_1d}
\end{figure*}

\begin{figure*}
\hspace*{-0.75cm}
\begin{tabular}{*2{c}}
A) EXO 0748$-$676: Spectrum 8 & B) EXO 0748$-$676: Spectrum 18\\
\quad \\
\includegraphics[width=0.5\textwidth]{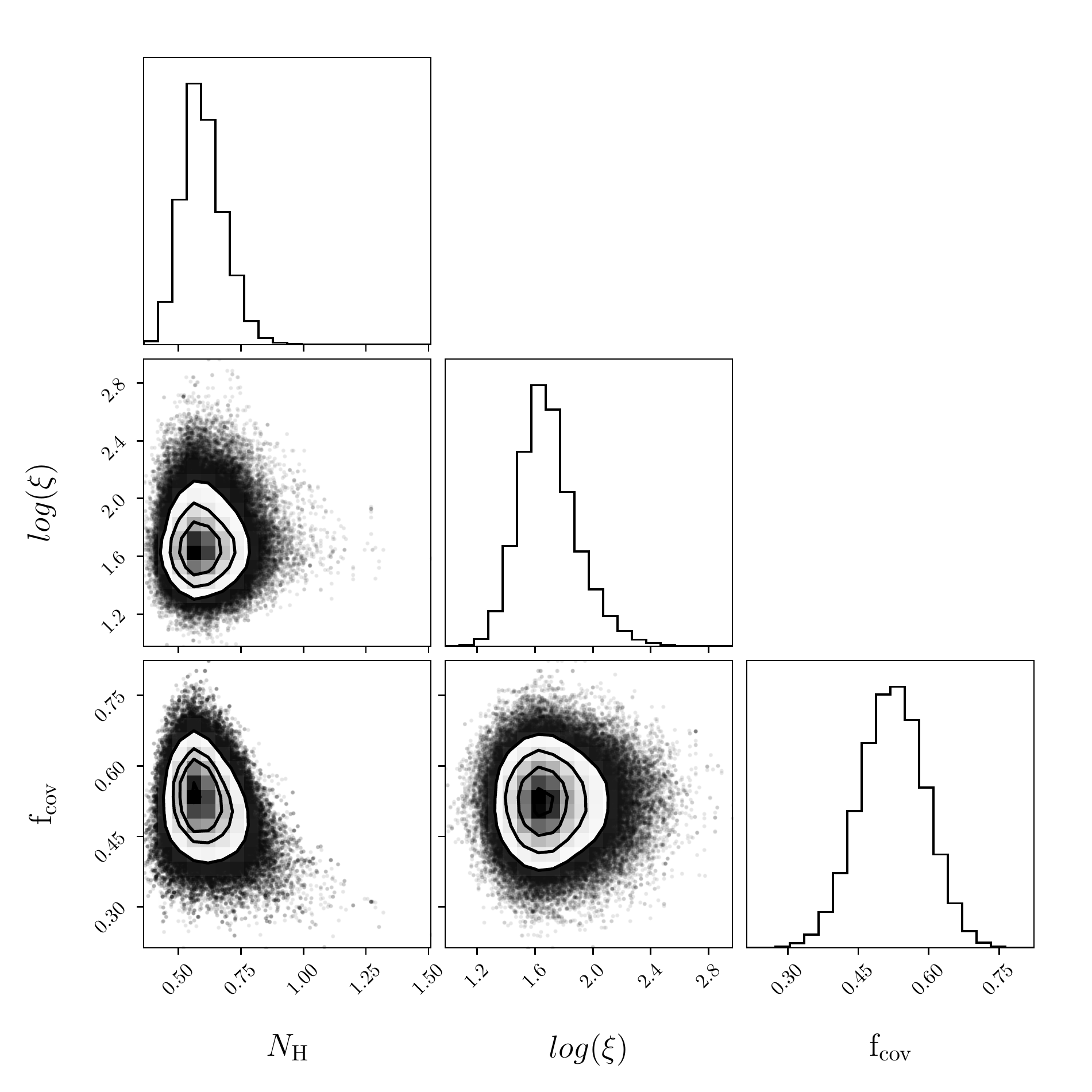}
&
\includegraphics[width=0.5\textwidth]{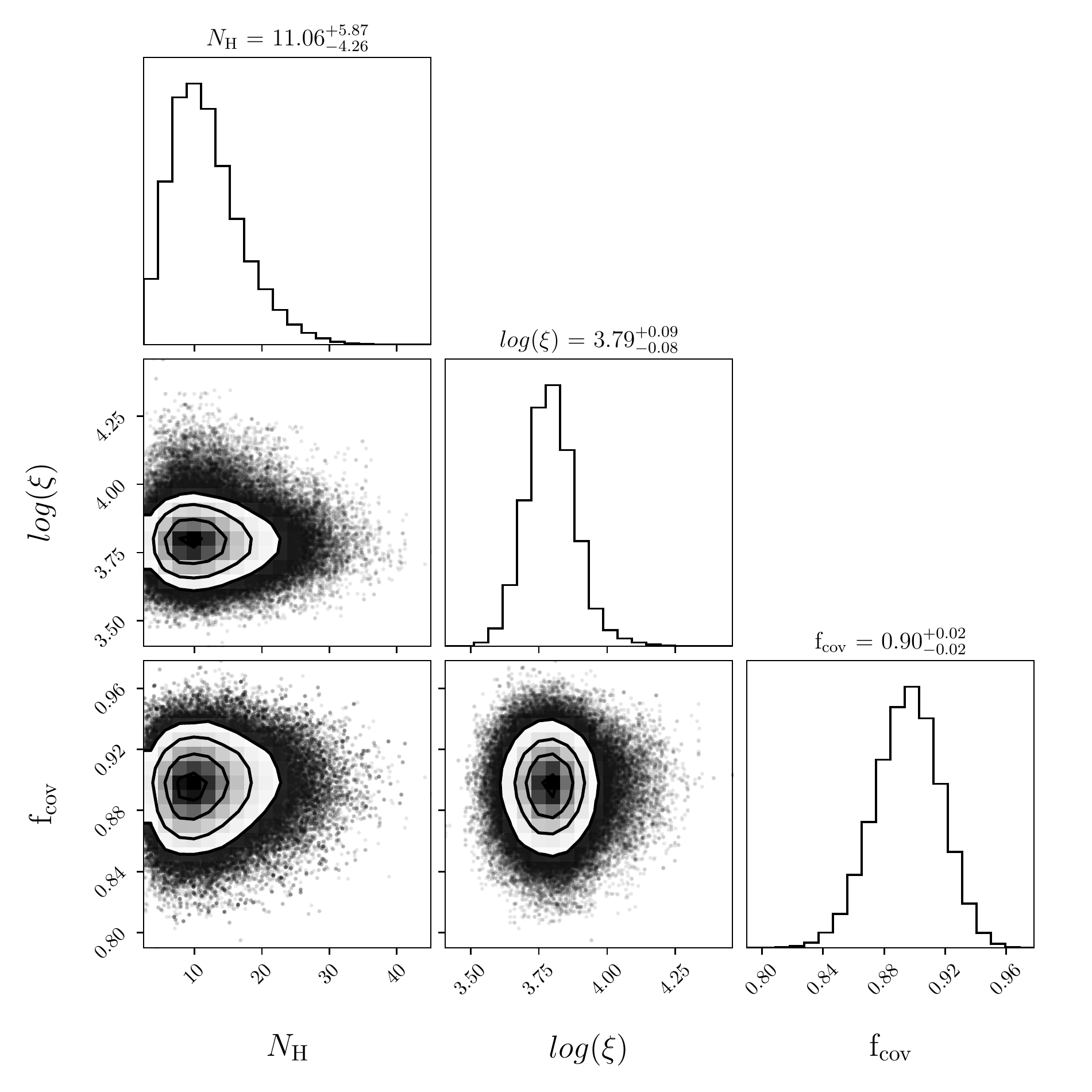} \\
\quad \\
\quad \\
C) EXO 0748$-$676: Spectrum 36 & D) EXO 0748$-$676: Spectrum 44\\
\quad \\
\includegraphics[width=0.5\textwidth]{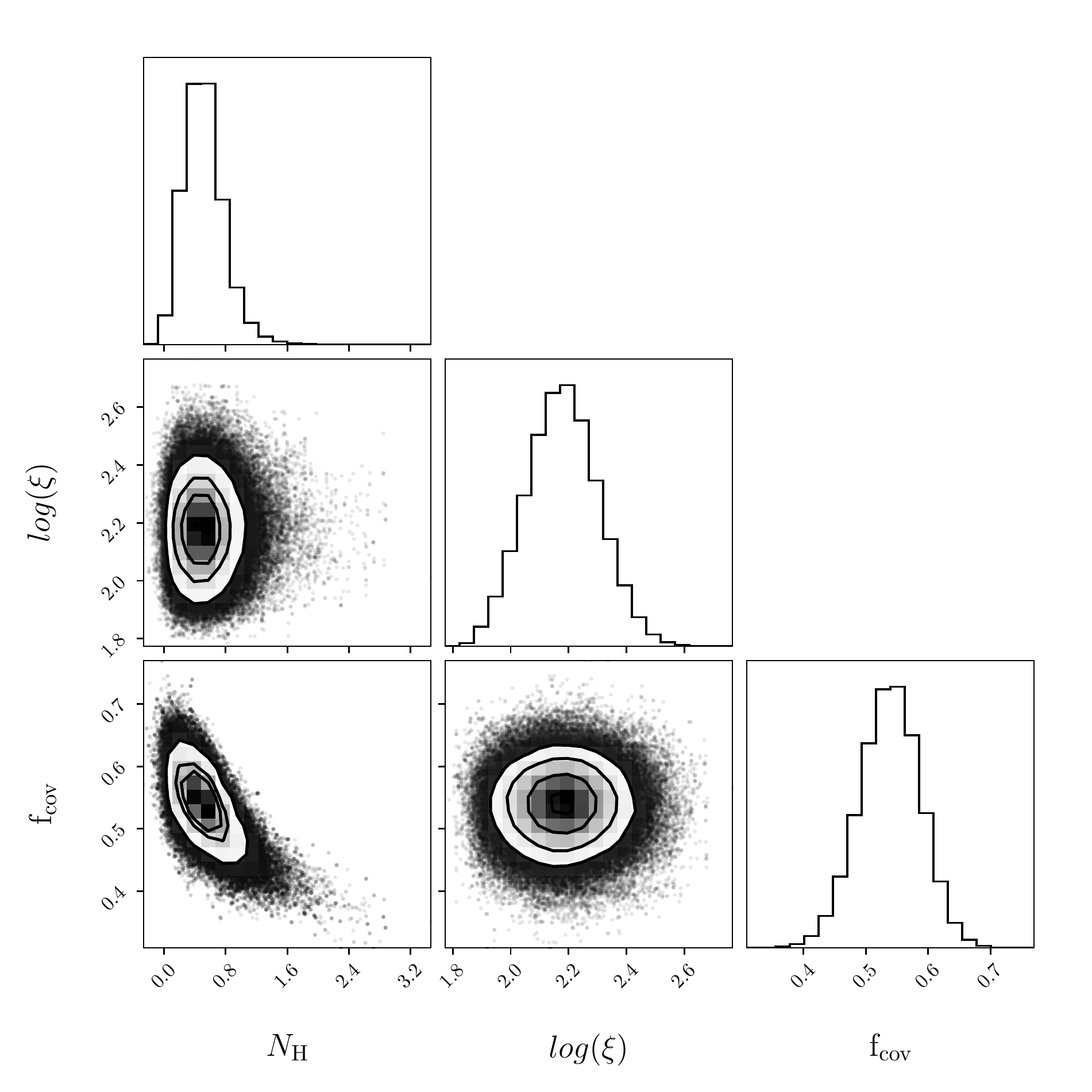}
&
\includegraphics[width=0.5\textwidth]{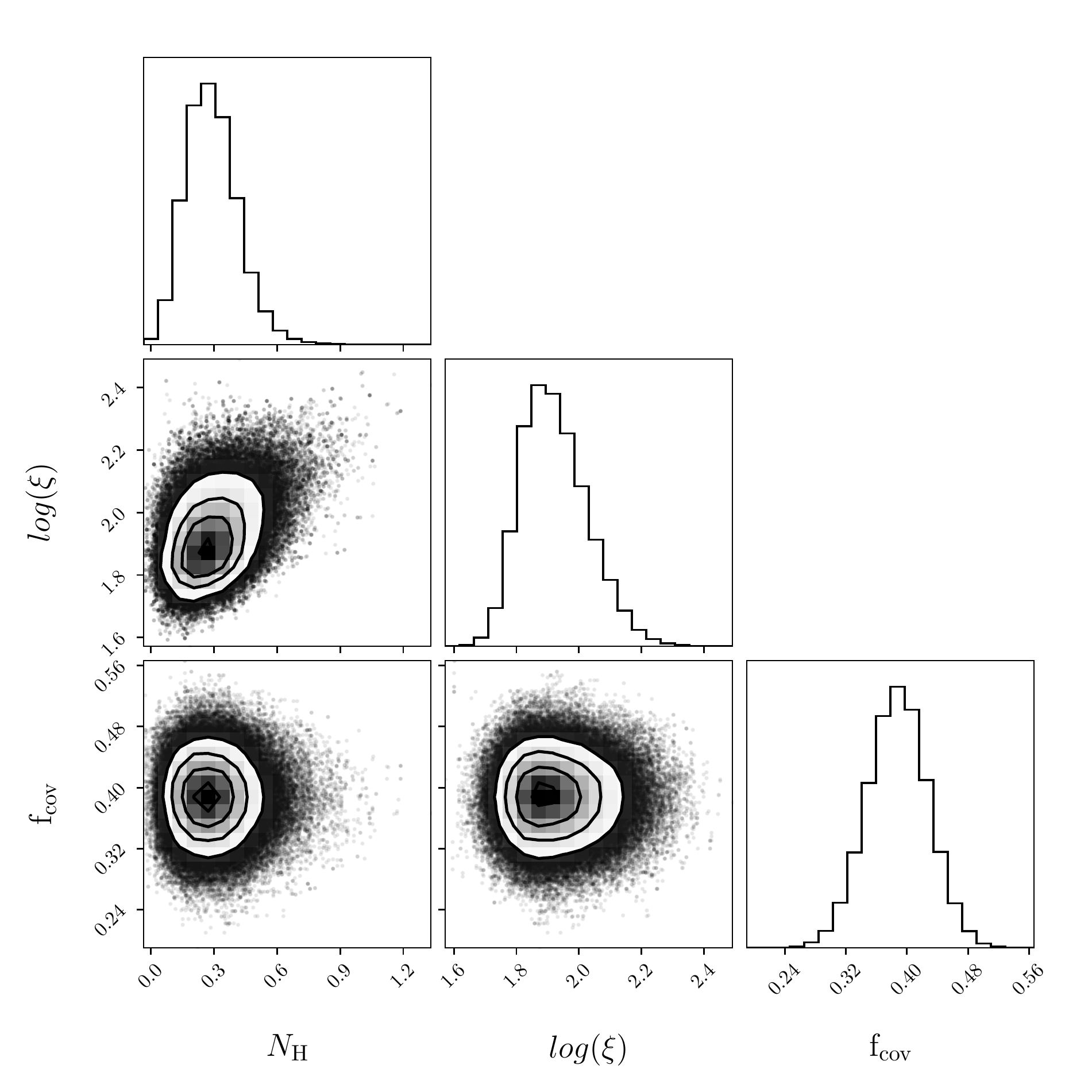} \\
\end{tabular}
\caption{Distributions of $N_{\rm{H}}$, $\log(\xi)$ and f$_{\rm{cov}}$ for four different phase-resolved spectra of EXO 0748 obtained by running an MCMC simulation of \textsc{abssca} within \textsc{xspec}. Spectra are numbered following a left-to-right convention of the spectral fit in Fig. \ref{fig:EXO_PRS}. The chains are run using the Goodman-Weare algorithm, using a length of 307200, 256 walkers and a burn-in period of 19998. For the 2D histograms, $1 \sigma$, $2 \sigma$ and $3 \sigma$ contours are shown by the solid black lines. The 1D histograms are displayed with their y-axes in arbitrary units.}
\label{fig:EXO2D}
\end{figure*}

\begin{figure*}
\centering
\includegraphics[width=\textwidth]{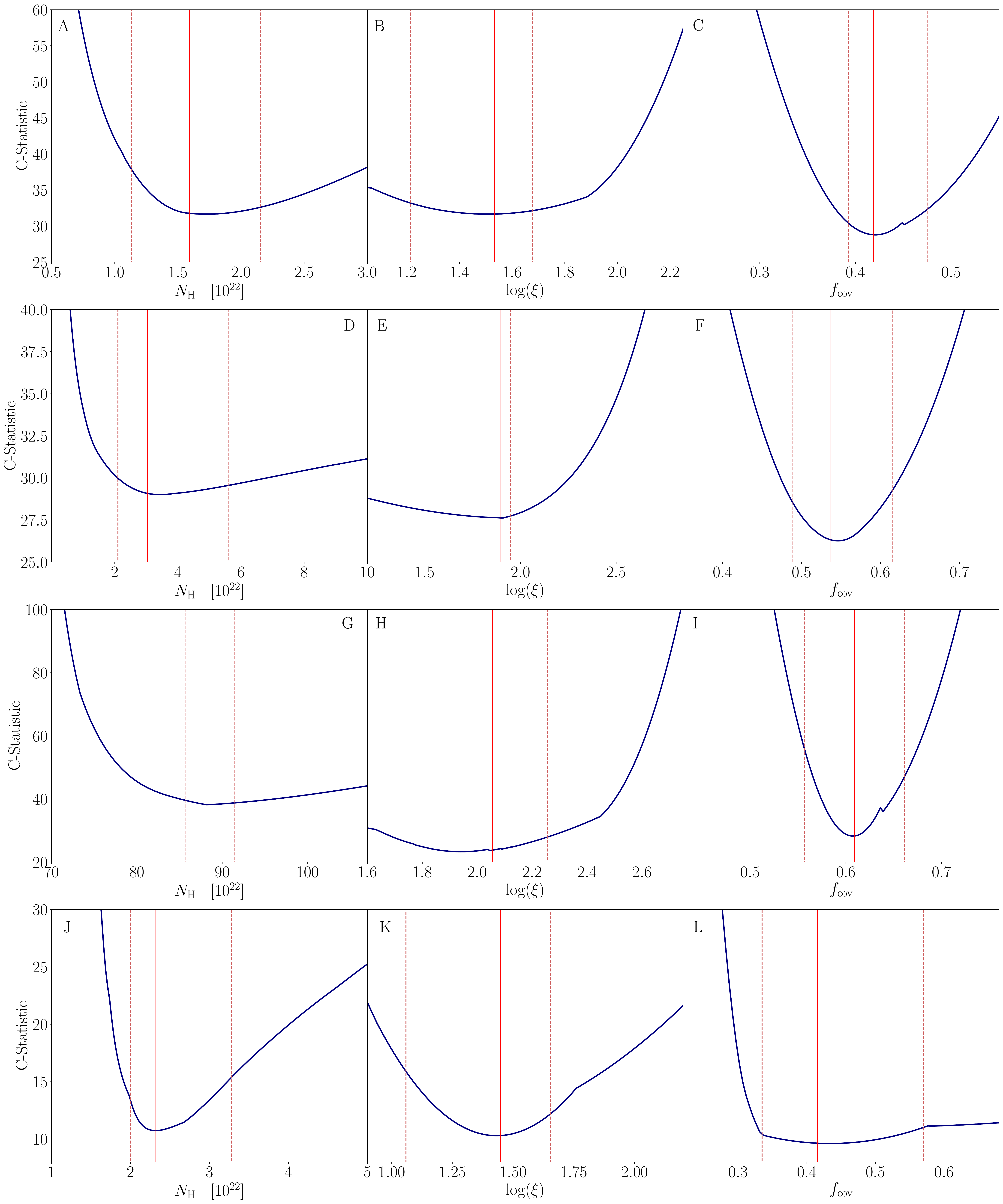}
\vspace{-0.5cm}
\caption{1D parameter contours (blue) for four different phase-resolved spectra of Sw J1858 obtained by running \texttt{steppar} analysis within \textsc{xspec}. Respectively, panels A-C, D-F, G-I and J-L correspond to Spectra 6, 13, 30 and 37, following a left to right numbering convention for the spectral fits in Fig. \ref{fig:J1858_PRS}. For each spectra $\log(\xi)$, f$_{\rm{cov}}$ were respectively varied in the ranges $0.75 - 3.0 $ and $0.05 - 0.95$ using 400 steps. For each spectra, $N_{\rm{H}}$ is varied through a range of values surrounding the best fitting values, again using 400 steps. For panels A this range is $1.0 \times 10^{-5} - 5.0$. For panel D and J the range used is $0.1 - 30.0$ and for panel G the range used is $50.0 - 200.0$. In each panel the best fitting parameter value is shown by the solid red line while the dashed red lines represent $1 \sigma$ contours. 
}
\label{fig:J10_1d}
\end{figure*}

\begin{figure*}
\hspace*{-0.75cm}
\begin{tabular}{*2{c}}
A) \textit{Swift} J1858.6$-$0814: Spectrum 6 & B) \textit{Swift} J1858.6$-$0814: Spectrum 13\\
\quad \\
\includegraphics[width=0.5\textwidth]{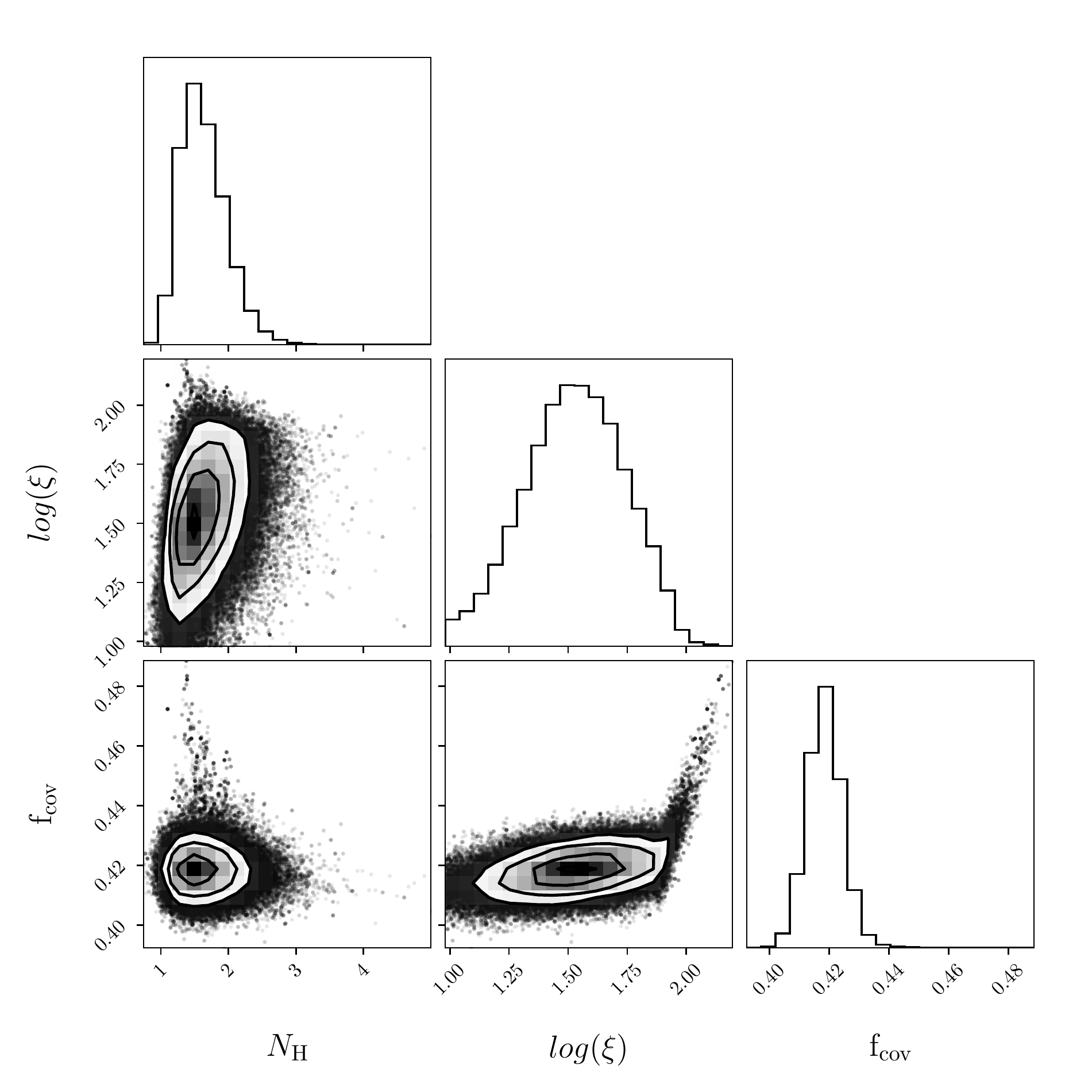}
&
\includegraphics[width=0.5\textwidth]{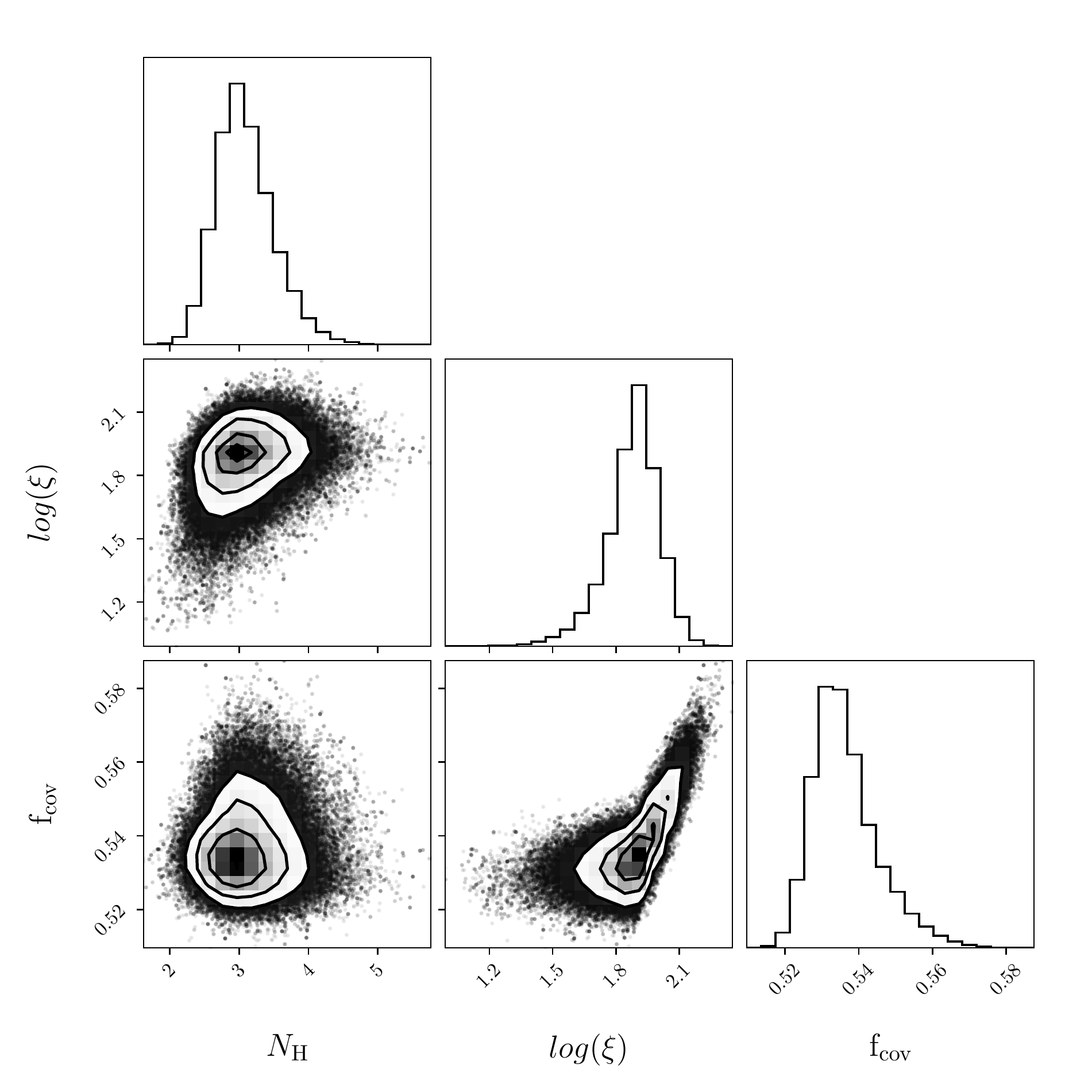} \\
\quad \\
\quad \\
C) \textit{Swift} J1858.6$-$0814: Spectrum 30 & D) \textit{Swift} J1858.6$-$0814: Spectrum 37\\
\quad \\
\includegraphics[width=0.5\textwidth]{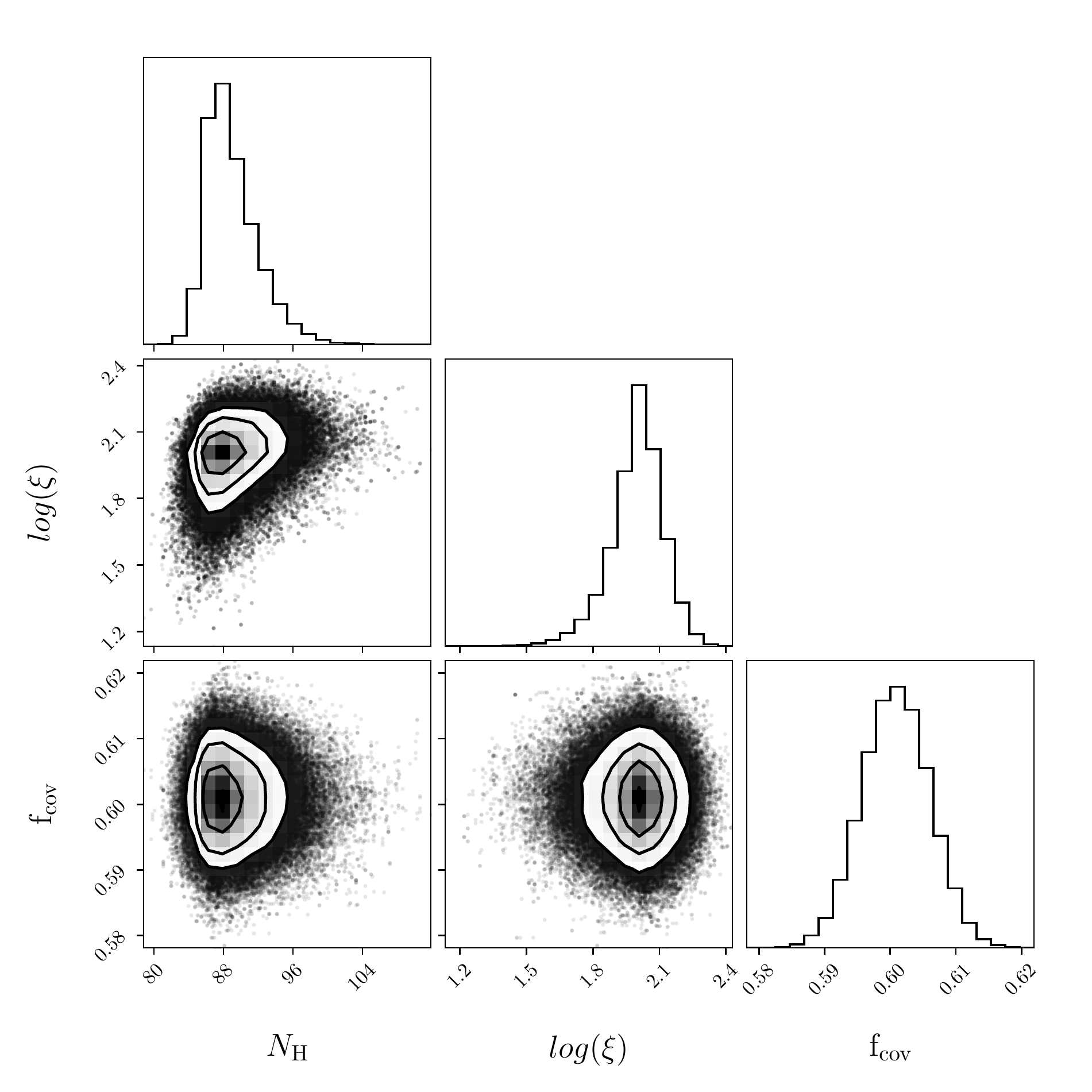}
&
\includegraphics[width=0.5\textwidth]{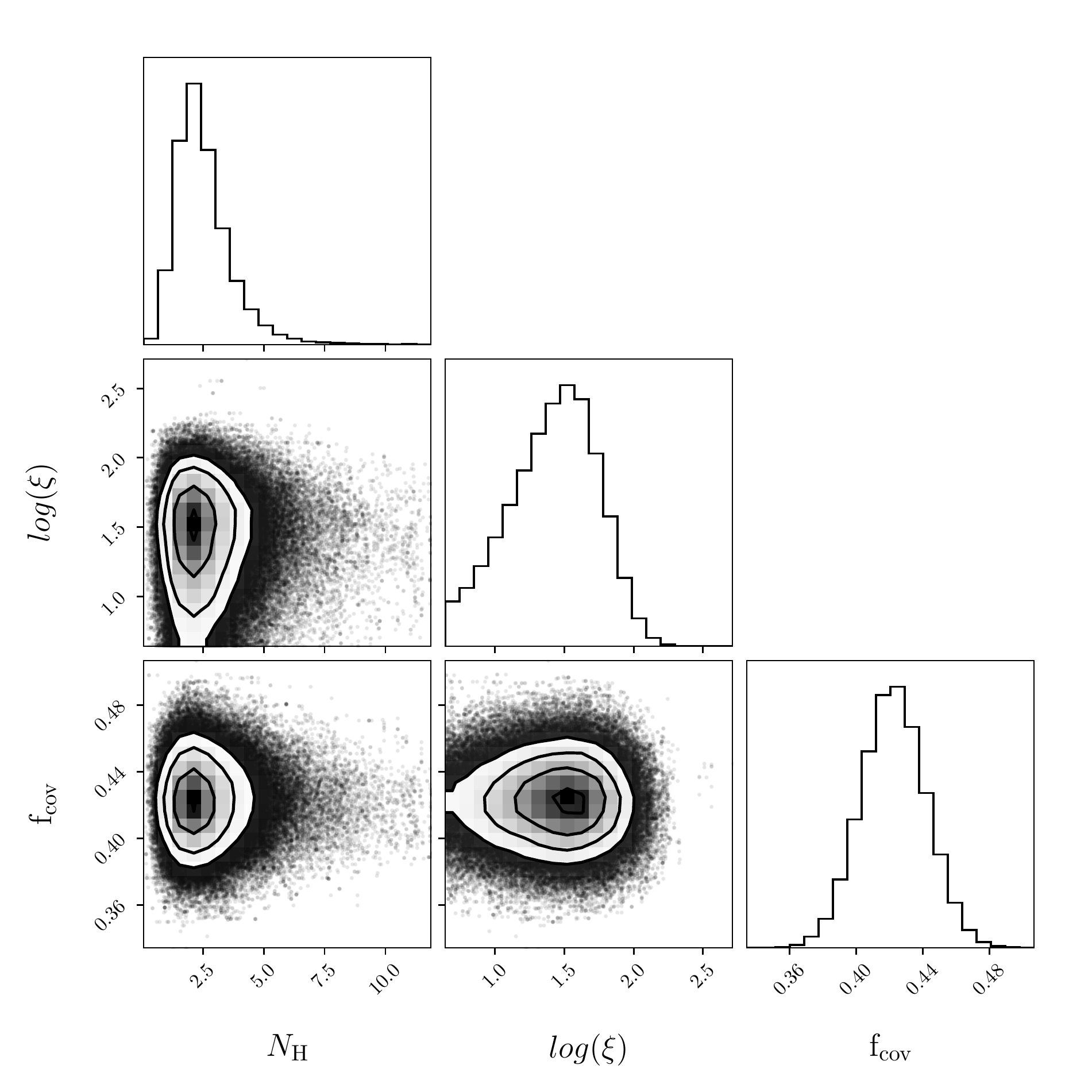} \\
\end{tabular}
\caption{Distributions of $N_{\rm{H}}$, $\log(\xi)$ and f$_{\rm{cov}}$ for four different phase-resolved spectra of Sw J1858 obtained by running an MCMC simulation of \textsc{abssca} within \textsc{xspec}. Spectra are numbered following a left-to-right convention of the spectral fit in Fig. \ref{fig:J1858_PRS}. The chains are run using the Goodman-Weare algorithm, using a length of 307200, 256 walkers and a burn-in period of 19998. For the 2D histograms, $1 \sigma$, $2 \sigma$ and $3 \sigma$ contours are shown by the solid black lines. The 1D histograms are displayed with their y-axes in arbitrary units.}
\label{fig:J2D}
\end{figure*}

For each representative spectrum and each parameter, we perform a 1-dimensional \texttt{steppar} analysis within \textsc{xspec} to check how well each parameter is constrained and whether a better fit (one with an overall lower C-statistic) could be found. The results of each \texttt{steppar} are shown by the 1-dimensional contour plots in Figs. \ref{fig:ex10_1d} and \ref{fig:J10_1d} respectively for EXO 0748 and Sw J1858. In these plots, the solid red line highlights the best-fitting parameter value and the dashed red lines indicate the $1 \sigma$ confidence interval. Here, we clearly see that each parameter \textit{can} be individually constrained in each of the representative spectra, although in some cases this is a broad constraint. While these contours are informative, they do not necessarily prove that partial parameter degeneracies have been avoided, nor do they highlight any parameter correlations. Therefore, for each representative spectra, we also include a corner plot showing the parameter distributions obtained by running a Markov-Chain Monte-Carlo (MCMC) simulation within \textsc{xspec}, using the Goodman-Weare algorithm. We use a chain length of 307200, 256 walkers and a burn-in period of 19998 to produce the corner plots in Figs. \ref{fig:EXO2D} and \ref{fig:J2D} respectively for the representative spectra of EXO 0748 and Sw J1858. These 2-dimensional parameter distributions largely support the conclusion from our 1-dimensional contours that each parameter can be individually constrained for each spectrum. Note that individually constraining each parameter was only possible when performing the analysis using c-stat, as is appropriate for the quality of our spectra. No strong parameter correlations are present, although in a few spectra tentative correlations do appear; e.g. Fig \ref{fig:EXO2D}C. If such correlations appeared in all corner plots it would be obvious that some parameter degeneracies were impacting our results, however, they are not a consistent feature. Therefore, we remain confident that each parameter in our modelling can be constrained in each spectrum and that the behaviours found in Figs. \ref{fig:EXO_PRS} and \ref{fig:J1858_PRS} are real.

Finally, we ensure that our choice to keep the data binned in six wide energy bands, thus maintaining consistency between this analyses and our earlier investigations of the same sources, does not impact our findings. For each of the eight representative spectra, we rebin the data to have one count per bin and repeat the fitting procedure. While the data do not necessarily need to be binned at all when modelling with c-stat, unbinned data are known to show biases that are not present when the data are finely binned \citep{Arnaud1996}. We compare the results of the finely binned spectra to those of the heavily binned spectra used in Section \ref{sx:PRS} in Table \ref{tb:cstatvschi}, finding that the best fitting parameters are generally consistent within $1 \sigma$ confidence intervals for the two binning procedures. While our choice to heavily bin the data does not appear to have influenced our findings, we note that more precise results are possible with finer binned spectra.

\begin{footnotesize}
\onecolumn
\setlength\LTcapwidth{\textwidth}
\begin{longtable}{p{2.18cm} p{0.35cm} p{1.65cm} p{1.65cm} p{1.65cm} p{1.85cm} p{0.45cm} lp{0.78cm} lp{0cm}}
\caption{\label{tb:RXTE_Table}Eclipse transition durations and X-ray hardness from all archival \textit{RXTE} observations of EXO 0748$-$676 containing full eclipse profiles. The full version of this table is available online.}\\
\toprule
Obs ID & PCU & Mid-Eclipse & Ingress & Egress & Totality & $\chi^{2}_{\nu}$ & $p$ & Rate & Hardness \\
& No. & MJD & Duration [s] & Duration [s] & Duration [s] & & & [cts/s] & $\frac{F_{10-16 \rm{keV}}}{F_{6-10 \rm{keV}}}$\\
\midrule
10108-01-01-00 & 5 & 50206.37478 & 1.996 $\pm$ 0.149 & 3.179 $\pm$ 0.238 & 508.405 $\pm$ 0.508 & 1.024 & 0.578 & 150.3 & 0.730 \\
10108-01-02-00 & 5 & 50206.53410 & 1.983 $\pm$ 0.149 & 3.055 $\pm$ 0.229 & 505.025 $\pm$ 0.505 & 1.001 & 0.682 & 160.6 & 0.788 \\
10108-01-03-00 & 5 & 50206.69343 & 3.073 $\pm$ 0.231 & 4.455 $\pm$ 0.334 & 502.006 $\pm$ 0.602 & 1.025 & 0.573 & 129.8 & 0.742 \\
10108-01-04-00 & 5 & 50206.85277 & 3.070 $\pm$ 0.230 & 4.514 $\pm$ 0.338 & 504.060 $\pm$ 0.604 & 1.014 & 0.606 & 136.6 & 0.713 \\
10108-01-05-00 & 5 & 50207.01211 & 1.984 $\pm$ 0.148 & 3.141 $\pm$ 0.236 & 504.306 $\pm$ 0.647 & 1.005 & 0.643 & 136.9 & 0.749 \\
10068-03-02-00 & 5 & 50298.15333 & 3.016 $\pm$ 0.226 & 5.666 $\pm$ 0.424 & 498.558 $\pm$ 0.257 & 1.000 & 0.688 & 153.3 & 0.894 \\
10108-01-06-00 & 5 & 50310.10367 & 2.958 $\pm$ 0.222 & 9.497 $\pm$ 0.712 & 492.059 $\pm$ 0.591 & 1.000 & 0.699 & 139.4 & 0.846 \\
10108-01-07-00 & 5 & 50310.26303 & 7.611 $\pm$ 0.571 & 4.057 $\pm$ 0.304 & 493.005 $\pm$ 0.391 & 1.019 & 0.594 & 135.4 & 0.876 \\
10108-01-08-00 & 5 & 50310.42237 & 6.177 $\pm$ 0.463 & 5.612 $\pm$ 0.421 & 494.553 $\pm$ 0.568 & 1.001 & 0.682 & 134.0 & 0.786 \\
10108-01-09-00 & 4 & 50310.58170 & 5.031 $\pm$ 0.227 & 9.098 $\pm$ 0.682 & 494.551 $\pm$ 0.566 & 1.011 & 0.615 & 82.84 & 0.795 \\
10108-01-07-01 & 5 & 50310.90039 & 6.481 $\pm$ 0.486 & 10.92 $\pm$ 0.819 & 492.593 $\pm$ 0.395 & 1.003 & 0.659 & 155.3 & 0.522 \\
\hline
\end{longtable}
\end{footnotesize}

\setlength\LTcapwidth{\textwidth}
\begin{longtable}{p{1.8cm} p{2.0cm} p{2.0cm} p{2.0cm} p{2.0cm} p{2.0cm} p{2.0cm}}
\caption{\label{tb:cstatvschi} $1 \sigma$ confidence intervals for each of the three key \textsc{abssca} model parameters; the material column density N$_{\rm{H}}$, the log of the ionisation parameter, log($\xi$) and the material covering fraction, f$_{\rm{cov}}$, for each of the eight representative spectra. We compare these intervals for the cases where the spectra are heavily (h) binned into the six energy bands (see Section \ref{sx:PRS}) and finely (f) binned to have one count per bin. The parameters are found to be mostly consistent within $1\sigma$ confidence bounds. All fits were carried out using c-stat.}\\
\toprule
Spectra & N$_{\rm{H}}$ [$10^{22}$] (h) & N$_{\rm{H}}$ [$10^{22}$] (f) & log($\xi$) (h)  & log($\xi$) (f) & f$_{\rm{cov}}$ (h) & f$_{\rm{cov}}$ (f) \\
\midrule
EXO 0748: 8 & 0.4252 - 0.8167 & 0.5907 - 0.6020 & 1.5456 - 1.8013 & 1.7407 - 1.7572 & 0.4379 - 0.6494 & 0.5323 - 0.5359 \\
EXO 0748: 18 & 7.2929 - 21.016 & 11.659 - 13.209 & 3.6398 - 3.9675 & 4.2809 - 4.3117 & 0.8272 - 1.000 & 0.8852 - 0.9414 \\
EXO 0748: 36 & 0.3721 - 0.6823 & 0.5176 -  0.5309 & 2.0941 - 2.2819 & 2.1958 - 2.2080 & 0.4963 - 0.5954 & 0.5029 - 0.5330 \\
EXO 0748: 44 & 0.1452 - 0.4820 & 0.4969 - 0.5649 & 1.6512 - 2.4598 & 1.4883 - 1.5697 & 0.3537 - 0.4326 & 0.3488 - 0.3722 \\
\midrule
Sw J1858: 6 & 1.1347 - 2.1545 & 1.6074 - 1.6609 & 1.2148 - 1.6768 & 1.5521 - 1.6106 & 0.3931 - 0.4749 & 0.4149 - 0.4249 \\
Sw J1858: 13 & 2.1031 - 5.6160 & 3.0193 - 3.0544 & 1.7689 - 1.9189 & 1.8774 - 1.8929 & 0.4889 - 0.6157 & 0.5309 - 0.5354 \\
Sw J1858: 30 & 85.744 - 91.901 & 90.251 - 92.546 & 1.6461 -  2.2557 & 2.0730 - 2.0899 & 0.5571 - 0.6612 & 0.6085 - 0.6089 \\
Sw J1858: 37 & 1.9995 - 3.2778 & 2.9559 - 2.9979 & 1.0589 - 1.6548 & 1.3049 - 1.3211 & 0.3349 - 0.5705 & 0.3493 - 0.3509 \\
\toprule
\end{longtable}

\end{appendices}


\bsp	
\label{lastpage}
\end{document}